\documentclass[preprint,12pt]{elsarticle}
\usepackage{graphicx} 
\usepackage{amssymb}
\usepackage{amsmath}
\usepackage{xcolor}
\usepackage{bm}
\usepackage{float}
\usepackage{makecell}
\usepackage{url}
\usepackage{hyperref}
\usepackage{threeparttable}
\usepackage{booktabs}
\usepackage{longtable}
\usepackage{titlesec}
\usepackage{nameref}
\usepackage{enumitem}
\usepackage{algorithm}
\usepackage{algpseudocode}
\usepackage{gensymb}
\usepackage{array} 
\usepackage{multirow}
\usepackage{rotating}

\renewcommand{\thesubsubsection}{\alph{subsubsection})}

\titleformat{\subsubsection}[runin] 
  {\bfseries} 
  {\thesubsubsection} 
  {3pt} 
  {}

\titlespacing{\subsubsection}
  {0pt} 
  {0pt} 
  {3pt}

\DeclareMathOperator*{\argmin}{arg\,min}

\journal{International Journal of Electrical Power and Energy Systems}

\begin{document}

\begin{frontmatter}

\title{Probabilistic Assessment of Rare Transient Instability Events via Kriging-based Active Learning Framework}

\author[mcgill]{Jingyu Liu}
\ead{jingyu.liu@mail.mcgill.ca}

\author[ualberta]{Xiaoting Wang}
\ead{xiaotin5@ualberta.ca}

\author[mcgill]{Xiaozhe Wang\corref{cor1}}
\ead{xiaozhe.wang2@mcgill.ca}

\cortext[cor1]{Corresponding author}

\affiliation[mcgill]{
    organization={Department of Electrical and Computer Engineering, McGill University},
    city={Montréal},
    state={QC},
    postcode={H3A 0E9},
    country={Canada}
}

\affiliation[ualberta]{
    organization={Department of Electrical and Computer Engineering, University of Alberta},
    city={Edmonton},
    state={AB},
    postcode={T6G 1H9},
    country={Canada}
}

\begin{abstract}

    The increasing uncertainty in modern power systems, driven by the integration of intermittent energy sources and variable loads, underscores the need for probabilistic transient stability assessment.  
    However, existing assessment methods
    primarily focus on average system stability behavior and may struggle or incur high computational cost when identifying rare transient instability events, which in turn are critical for ensuring system resilience. 
    To address this, the paper proposes a Kriging-based active learning framework
    to accurately characterize rare instability regions within the input uncertainty space and estimate the associated small instability probability, 
    while requiring only a limited number of expensive time-domain simulations.
    \color{black}
    The proposed active learning (AL) framework is tested on a modified IEEE 59-bus system with simulated load and wind uncertainties, and a WECC 240-bus system incorporating real-world wind and solar generation data.
    \color{black}
    Comparative studies with the existing random forest-based active learning method and 
    three non-AL methods demonstrate that the proposed AL framework achieves superior accuracy and computational efficiency. 
\end{abstract}

\begin{keyword}
Active learning, Kriging, critical clearing time (CCT),  probabilistic transient stability, uncertainty quantification, surrogate model, rare events.
\end{keyword}

\end{frontmatter}

\section{Introduction}


With the increasing integration of renewable energy sources (RES) and the growing variability of load demand, modern power systems exhibit more complex dynamic behavior and operate under increasingly uncertain conditions.
\color{black}
Consequently, transient stability assessment (TSA) has become increasingly challenging. 
Traditional TSA methods, such as time-domain simulation (TDS) and direct methods \cite{chiang2011direct}, are typically conducted on a limited number of representative operating scenarios \cite{nerc2014probabilistic}. However, this representative-case-based paradigm is becoming insufficient for modern power systems, as it cannot adequately capture the variability of operating conditions or reflect their likelihood of occurrence.

To address this challenge, recent research has evolved along two complementary directions. 
The first is deterministic online TSA \cite{sarajcev2022artificial}, which employs offline-trained machine learning (ML) models to achieve fast online screening of system stability under diverse operating conditions. In this paradigm \cite{tan2019deep, zhan2023hybrid, liu2026two}, the ML model is trained offline on a large labeled dataset, often generated from TDS, and then applied online using PMU measurements for case-wise stability prediction.

The second is planning-oriented probabilistic TSA \cite{papadopoulos2016probabilistic}, which models uncertainties in operating conditions as random variables with associated probability distributions and assesses system stability probabilistically through statistical sampling. 
\color{black}The most accurate probabilistic TSA method is the time-domain simulation-based Monte Carlo simulation (TDS-MCS) \cite{papadopoulos2016probabilistic}, which estimates probabilistic stability metrics by performing a large number of TDS over samples drawn from the input probability distributions. However, repetitive TDS can be computationally intensive and impractical for large-scale systems. 
To improve efficiency, ML-based methods have been proposed 
\cite{tan2023debiased, lu2024advanced, xu2018propagating, ye2022physics,liu2022sparse}. Unlike ML-based online TSA methods
\cite{tan2019deep, zhan2023hybrid, liu2026two}, these methods are designed to estimate probabilistic stability metrics, such as the probability of transient instability or the probabilistic distribution of a stability margin (e.g., critical clearing time),  using only a limited number of TDS runs, rather than to provide accurate pointwise prediction of transient stability status from PMU data. 
\color{black}
The difference between online TSA and probabilistic TSA from a machine learning perspective is summarized in Table \ref{tab:TSA_compare}, which shows that the two paradigms serve complementary purposes rather than replacing one another.
\begin{table}[!h]
\centering
\footnotesize
\renewcommand{\arraystretch}{1.35}
\setlength{\tabcolsep}{4pt}
\caption{\textcolor{black}{Comparison of deterministic online TSA and probabilistic TSA from a machine learning perspective}}
\label{tab:TSA_compare}
{\color{black}
\begin{tabular}{>{\centering\arraybackslash}m{2.8cm}
                >{\centering\arraybackslash}m{4.9cm}
                >{\centering\arraybackslash}m{4.9cm}}
\toprule
\textbf{Aspect} & \textbf{Online TSA} & \textbf{Probabilistic TSA} \\
\midrule
Input features & PMU-based dynamic measurements & Uncertain operating-condition variables (e.g., wind speed)\\
\addlinespace[2pt]
Input uncertainty modeling & Not explicitly required & Explicitly required\\
\addlinespace[2pt]
Role of statistical sampling & Not fundamentally required & Essential for uncertainty propagation \\
\addlinespace[2pt]
Training set size & Typically large, often $>10^4$ & Usually smaller, often $10^2$--$10^3$\\
\addlinespace[2pt]
Typical model output & \makecell[c]{Stability status\\ (i.e., stable or unstable)} & \makecell[c]{Stability margin \\(e.g., critical clearing time)}\\
\addlinespace[2pt]
Performance focus & Case-wise stability status prediction accuracy & Estimation accuracy of probabilistic stability metrics (e.g., instability probability)\\
\addlinespace[2pt]
Evaluation objective & Fast and accurate pointwise stability prediction & Efficient probabilistic stability characterization\\
\addlinespace[2pt]
Typical application & Online screening and emergency support & Risk assessment and preventive control\\
\bottomrule
\end{tabular}
}
\end{table}

\color{black}
This work focuses on the probabilistic TSA problem, for which several ML-based approaches have been developed in previous studies. Deep neural network (DNN)-based methods, such as multilayer perceptron (MLP) \cite{tan2023debiased} and graph neural network (GNN) \cite{lu2024advanced}, have been developed to estimate probabilistic transient stability metrics, such as the distribution of transient stability index and the probability of instability. Despite their stronger nonlinear modeling capability, these DNN-based methods generally require large training datasets (e.g., \(>10^3\) samples) and high-dimensional input features.
Alternatively, surrogate model-based methods such as polynomial chaos expansion (PCE) \cite{xu2018propagating,liu2022sparse} and Gaussian process regression (i.e., Kriging) \cite{ye2022physics} are generally more sample-efficient than DNN-based methods and are therefore more suitable for probabilistic TSA under limited TDS budgets. PCE \cite{xu2018propagating,liu2022sparse} can efficiently quantify the uncertainty of system responses by approximating the dynamic model and extracting output statistics from the expansion coefficients. However, conventional PCE typically requires prior knowledge of the input distributions, and its accuracy may deteriorate if correlations among uncertain inputs are not properly handled. A recent data-driven PCE method (DDPCE) \cite{wang2020data} eliminates this requirement, but its effectiveness for probabilistic TSA remains unverified. By contrast, Gaussian process regression (i.e., Kriging) \cite{ye2022physics} does not require prior knowledge of the input distributions and is suitable for small-sample probabilistic modeling.
\color{black}

\color{black}

\color{black}

 Despite these achievements, previous studies \cite{papadopoulos2016probabilistic, tan2023debiased, lu2024advanced, xu2018propagating, ye2022physics, liu2022sparse} primarily focus on average transient stability behaviours—such as the mean, variance, and probability density function of stability margins—typically under conditions where transient instability probability is relatively high (e.g., a 10\%-30\% probability).
  \color{black}
 In practice, however, transient instability events are often rare and may occur with much lower probabilities. While the overall stability of modern power systems may continue to improve with advances in system operation and control, accurately identifying low-probability but high-impact unstable scenarios remains crucial for ensuring system resilience.
 Existing probabilistic TSA methods may fail to capture unstable scenarios in the extreme tail regions (e.g., $<1\%$ probability)  due to the scarcity of unstable samples in training datasets.

To address this gap, we propose a Kriging-based active learning framework (AL-Kriging) for probabilistic transient instability assessment, 
with an explicit focus on 
identifying the low-probability instability regions within the space of uncertainties.
\color{black}
Active learning (AL) 
iteratively refines a surrogate model by adaptively selecting and sequentially enriching new training samples based on 
the predictions of the current surrogate model and a specified objective, 
such as more accurate characterization of the stability boundary.
\color{black}
While AL has been applied in 
power system studies to reduce the training cost of a random forest-based stability classifier \cite{zhang2021power}, existing approaches have not effectively addressed rare event identification. In particular, entropy-based learning strategies used in \cite{zhang2021power} may fail to provide accurate assessments when the initial training dataset lacks unstable samples.  
The key contributions of this work are as follows:
\color{black}

\begin{itemize}
    \color{black}
    \item This work explicitly addresses low-probability extreme-tail instability events ($<1\%$ probability), which have received limited attention in previous probabilistic transient stability assessment studies. This problem is increasingly important for the secure and resilient operation of modern power systems, which may appear stable on average yet remain vulnerable to rare instability events under extreme combinations of operating conditions associated with multiple sources of uncertainty.
    \color{black}

    \item We develop a Kriging-based active learning framework (AL-Kriging) specifically tailored for characterizing rare unstable regions within the uncertainty space. Unlike the entropy-based learning function in \cite{zhang2021power}, the adopted U-learning function effectively guides the sampling and surrogate model training process toward the boundary of rare instability region, even when the initial training dataset contains no unstable samples. In addition, compared with the AL-Kriging method in \cite{moustapha2022active}, the proposed method incorporates a multi-sample enrichment strategy and a newly designed stopping criterion to reduce the computational cost of iterative model training.

    \item The proposed AL-Kriging method enables reliable identification of rare transient instability events with limited TDS budgets (hundreds of TDS runs). Its performance is evaluated against
    the existing random forest-based AL method \cite{zhang2021power} and three non-AL methods:  two surrogate models (DDPCE \cite{wang2020data} and standard Kriging \cite{xu2020probabilistic}), and a deep neural network-based method (MLP \cite{tan2023debiased}). Results show that the proposed method achieves consistently superior accuracy in identifying rare transient instability events, even outperforming MLP models trained on an order-of-magnitude larger dataset.

\end{itemize}

\color{black}
It should be noted that identifying rare transient instability samples is related to, but fundamentally different from, the data-imbalance problem studied in online TSA \cite{tan2019deep, zhan2023hybrid, liu2026two}. Data imbalance refers to the severe disparity between stable and unstable samples, which degrades the training of machine-learning classifiers. Existing correction techniques address this issue by enriching the training dataset with additional unstable samples, typically assuming that at least hundreds of labeled unstable cases are already available from extensive TDS runs. In contrast, the proposed AL-Kriging framework starts with a large unlabeled sample set 
and aims to identify rare unstable samples in this sample set using only a limited number of TDS evaluations. Rather than rebalancing a labeled dataset, it adaptively selects unlabeled samples for TDS evaluation to discover rare instability events. 

\color{black}

The remainder of this paper is organized as follows: Section \ref{sec:ProbForm} formulates the problem of probabilistic transient stability assessment, including the definitions of the 
transient stability margin (TSM) and the probability of transient instability. Section \ref{sec:general_framework} describes the general AL framework and its components. Section \ref{sec:algorithm} presents the implementation of the proposed AL-Kriging method for rare instability event detection. Section \ref{sec:case_study} presents the case study results, and  Section \ref{sec:conclusion} concludes the paper.

\section{Problem Formulation}
\label{sec:ProbForm}

The power system transient dynamics are typically modelled by a set of differential-algebraic equations \cite{liu2022sparse}:

\begin{equation}
\begin{aligned}
\dot{\mathbf{x}}_s &= \bm{f}(\mathbf{x}_s, \mathbf{y}, \bm{\lambda}, \mathbf{u})  \\
\bm{0} &= \bm{h}(\mathbf{x}_s, \mathbf{y}, \bm{\lambda}, \mathbf{u})
\label{eq:PSDAE}    
\end{aligned}
\end{equation}

%

\noindent where $\bm{f}$ are differential equations (e.g., swing equations); $\bm{h}$ are algebraic equations (e.g., power flow); $\mathbf{x}_s$ are state variables (e.g., rotor speeds and rotor angles of generators); $\mathbf{y}$ are algebraic variables (e.g., bus voltage magnitudes and angles); $\bm{\lambda}$ are system parameters (e.g., load power); $\mathbf{u}$ are discrete variables modeling events (e.g., fault occurrence and switching operation of circuit breakers).

\begin{figure}[!h]
 \vspace{-0.3cm}
\centering
\includegraphics[width=0.75\linewidth]{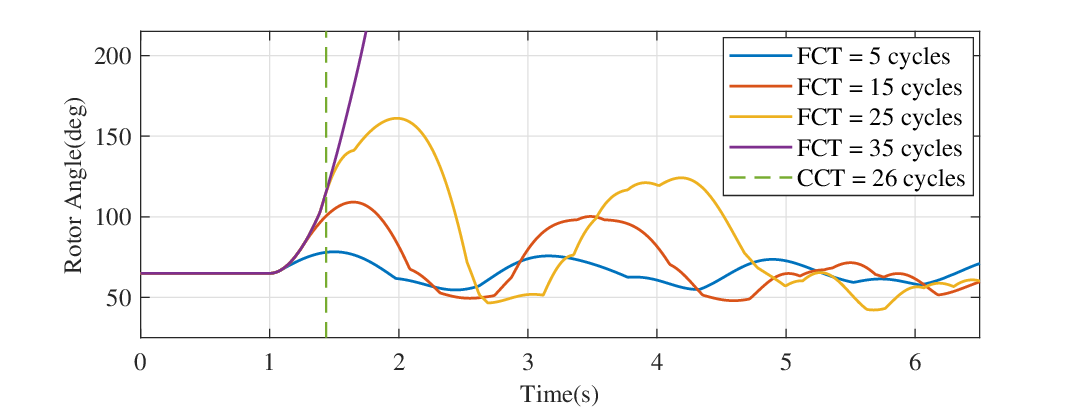}
\vspace{-0.3cm}
\caption{Illustration of maximum rotor angle difference trajectories. A fault is applied at $t = 1.0$s, and is cleared after different fault clearing times.\color{black}{ Here, 1 cycle = 1/60 s for the 60 Hz system.}}
\label{fig:CCTExplanation}
\vspace{-0.1cm}
\end{figure}

Numerical integration of \eqref{eq:PSDAE} yields the generator rotor angle trajectories. The time-varying maximum difference of these trajectories, as shown in Fig.~\ref{fig:CCTExplanation}, clearly indicates whether the system maintains transient stability.

In Fig. \ref{fig:CCTExplanation}, the fault clearing time (FCT, $T_{\mathrm{fct}}$) refers to the time duration between the fault occurrence ($t = 1.0$s) and its clearance. Increasing $T_{\mathrm{fct}}$ typically leads to instability (e.g., the purple solid line indicating diverged trajectories in Fig. \ref{fig:CCTExplanation}). The maximum value of $T_{\mathrm{fct}}$ that preserves system stability is defined as the critical clearing time (CCT, $T_{\mathrm{cct}}$). 
\color{black}
It is determined through a binary search procedure over $T_{\mathrm{fct}}$ values \cite{zhu2019hierarchical}.
\color{black} 
The difference $T_{\mathrm{cct}} - T_{\mathrm{fct}}$ is a well-accepted transient stability margin (TSM) in transient stability studies \cite{liu2022sparse}, indicating 
how close the system is to instability.


In deterministic cases, $T_{\mathrm{cct}}$ is a constant determined by fixed $\bm{\lambda}$ and $\bm{u}$. When the volatile RES are integrated into the system, parts of $\bm{\lambda}$ will become uncertain (e.g., wind and solar generation)  and can be modelled as random variables $\bm{X}$. Accordingly, \eqref{eq:PSDAE} can be recast as:

\begin{equation}
\begin{aligned}
         \dot{\mathbf{x}}_s &= \bm{f}(\mathbf{x}_s, \mathbf{y}, \bm{\lambda}^{\prime}, \bm{X},  \mathbf{u})  \\
           \bm{0} &= \bm{h}(\mathbf{x}_s, \mathbf{y}, \bm{\lambda}^{\prime}, \bm{X}, \mathbf{u})
\label{eq:Stochastic-PSDAE}        
\end{aligned}
\end{equation}

\noindent where $\bm{\lambda}^{\prime}$ denotes the parameters that remain constants. In this context, CCT also becomes a random variable, termed probabilistic CCT ($T_{\mathrm{pcct}}$), which leads to the formulation of the probabilistic TSM (PTSM):
\begin{equation}
     \mathrm{PTSM}:= \mathcal{M}(\bm{X}) = T_{\mathrm{pcct}}(\bm{X}) - T_{\mathrm{fct}}
    \label{eq:PCCT}
\end{equation}
Since $\mathcal{M}(\bm{X})$ cannot be expressed analytically, $T_{\mathrm{pcct}}$ is typically estimated via time-domain simulation (TDS) for a given realization of $\bm{X}$, 
which involves repeated numerical integration of \eqref{eq:Stochastic-PSDAE} and a bisection search over a range of fault clearing times.

An uncertain CCT may result in negative PTSM values, which in turn implies a non-zero probability $P_f$ that the system undergoes transient instability: 
\begin{equation}
    P_f = \mathbb{P}(T_{\mathrm{pcct}}(\bm{X})\leq T_{\mathrm{fct}}) = \mathbb{P}(\mathcal{M}(\bm{X})\leq 0)
\end{equation}
Assume that $\bm{X} \sim f_X(x)$, where $f_X(x)$ is the joint probability density function of $\bm{X}$, and define the domain of instability as $\mathcal{F} = \{\bm{x}: \mathcal{M}(\bm{x}) \leq 0\}$. The probability of transient instability is then given by:
\begin{equation}
    P_f = \int_{\mathcal{F}}f_{\bm{X}}(\bm{x})d\bm{x}
    \label{eq:Pf_calculate}
\end{equation}

The most common approach to solving \eqref{eq:Pf_calculate} is the TDS-based direct Monte Carlo simulation (TDS-MCS), 
whose results are often used as a reference benchmark due to their high accuracy. 
\color{black}
However, TDS-MCS is computationally intensive and particularly costly
for estimating small instability probabilities.
This is because direct Monte Carlo estimation requires a large number of TDS runs to adequately characterize the underlying low-probability instability region $\mathcal{F}$, since only a small fraction of samples will fall into $\mathcal{F}$. For example, when $P_f \approx 1\%$, TDS-MCS based on $10^5$ TDS runs is expected to yield around $10^3$ unstable samples, which is generally sufficient to characterize this rare instability region \cite{au2014SusBook}.
\color{black}
To improve efficiency, this paper proposes a Kriging-based active learning (AL) framework that characterizes the rare instability region and estimates the associated small $P_f$ using only a limited number of TDS evaluations. The Kriging model, refined through the AL process, enables more efficient identification of rare unstable samples than the existing random forest-based AL method \cite{zhang2021power} and non-AL methods (e.g., DDPCE \cite{wang2020data}). 
\color{black}


\section{Kriging-Based Active Learning Framework}
\label{sec:general_framework}

\begin{figure}[!ht]
    \centering 
    \includegraphics[width=1\linewidth]{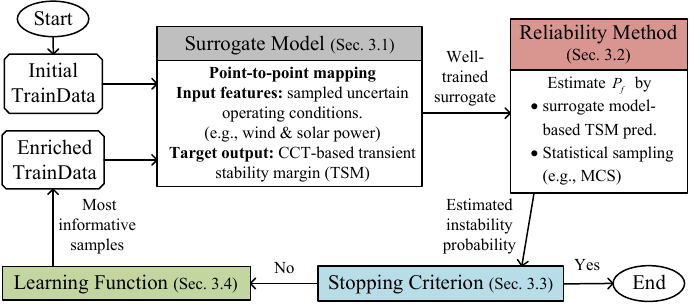}
    \caption{General framework of the active learning process for transient instability probability assessment. The active learning process is a loop starting with an initial training dataset (TrainData). The TrainData will be enriched with the most informative samples in each iteration to reach a more accurate prediction of the $P_f$. Our framework contains four elements: \textbf{a)} Kriging (Section \ref{subsec:Kriging_intro}) adopted as the surrogate model. \textbf{b)} MCS (Section \ref{subsec:Kriging_based_RA}) adopted as the reliability method. \textbf{c)} the stopping criterion (Section \ref{subssec:sc}). \textbf{d)} the learning function (Section \ref{subsec:LF}).}
    \label{fig:AL_explain}
    \vspace{-0.25cm}
\end{figure}

The general active learning framework \cite{moustapha2022active} consists of four main components, as illustrated in Fig.~\ref{fig:AL_explain}. Each is briefly introduced below.

\subsubsection{The Surrogate Model:} Let $\mathcal{X}=[\bm{x}^{(1)}, \dots, \bm{x}^{(N)}]^T$ be $N$ realizations of the input random variables $\bm{X}$, which represent uncertain load power, wind speed, and solar power in our study.
Let $\mathcal{T}=[\mathcal{M}(\bm{x}^{(1)}), \dots, \mathcal{M}(\bm{x}^{(N)})]^T$ denote the transient stability margins of $\mathcal{X}$ obtained via TDS. A surrogate model $\mathcal{M}_s(\bm{x})$, constructed based on the training dataset $(\mathcal{X}, \mathcal{T})$, approximates the true model of probabilistic transient stability margin $\mathcal{M}(\bm{x})$ defined in \eqref{eq:PCCT}:
\begin{equation}
    \mathcal{M}(\bm{x})\approx\mathcal{M}_s(\bm{x})
    \label{eq:surrogate_model_discription}
\end{equation}
$\mathcal{M}_s(\bm{x})$ is typically an explicit mapping that is significantly cheaper to evaluate than the TDS-based estimation of $\mathcal{M}(\bm{x})$. 
\color{black}
Note that although $\bm{X}$ and the probabilistic transient stability margin $\mathcal{M}(\bm{X})$ are random variables with associated probabilistic distributions, the surrogate model $\mathcal{M}_s(\bm{x})$ trained on their realizations $(\mathcal{X}, \mathcal{T})$ is a point-to-point mapping from a sample of $\bm{X}$ to its corresponding TSM values.
\color{black}
Possible choices of surrogate model training methods include Kriging \cite{xu2020probabilistic} and PCE \cite{wang2020data}.
In this work, we adopt Kriging due to its efficiency with small training datasets and its unique feature, the Kriging variance, which serves as a valuable metric for enriching the training dataset \cite{echard2011ak}.
Section \ref{subsec:Kriging_intro} details the principles of Kriging. 


\subsubsection{The Reliability Method:} Once the surrogate model is constructed, it is integrated with a reliability method to estimate the instability probability $P_f$.
Common approaches include MCS \cite{papadopoulos2016probabilistic}, importance sampling \cite{Dubourg2011}, and subset simulation \cite{au2014SusBook}. 
In this study, we adopt MCS as it can operate on raw data without requiring input distribution information, which is often unavailable in practice. Notably, the MCS here incurs negligible computational cost, as it is performed on the surrogate model $\mathcal{M}_s(\bm{x})$, rather than on the TDS-based estimation of $\mathcal{M}(\bm{x})$. 
See Section~\ref{subsec:Kriging_based_RA} for details.

\subsubsection{The Stopping Criterion:}
Following the surrogate model-based reliability analysis, 
the stopping criterion checks whether the predicted instability probability $\hat{P}_f$ has converged and terminates the algorithm accordingly.
See Section~\ref{subssec:sc} for details. 

\subsubsection{The Learning Function:} 
If the stopping criterion is not satisfied, the learning function will select the most informative samples (e.g., samples near the stability boundary) to enrich the training dataset for the next iteration of surrogate model training. 
In this work, we employ the U-learning function driven by Kriging predictive mean (i.e., the predicted stability margin) and variance, as detailed in Section~\ref{subsec:LF}.

\subsection{Kriging}
\label{subsec:Kriging_intro}
\begin{figure}[!ht]
    \centering  \includegraphics[width=0.7\linewidth]{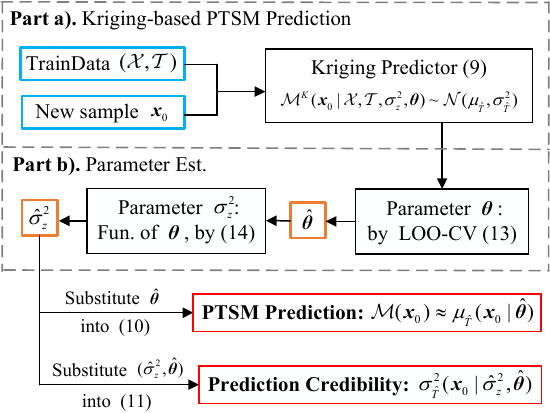}
    \caption{Overview of the Kriging method. The key results, the Kriging mean and variance, are highlighted in red. Their derivations are explained in \textbf{Part} \ref{subsubsec:Kriging_mean_var}; the required input data are outlined in blue. The Kriging model parameters are enclosed in orange. \textbf{Part} \ref{subsubsec:K_para_est} details the estimation of these parameters.}
    \label{fig:Kriging_explain}
    \vspace{-0.4 cm}
\end{figure}

The Kriging method uses a small training dataset $(\mathcal{X}, \mathcal{T})$ to build a surrogate model that predicts the probabilistic transient stability margin (PTSM) $\mathcal{M}(\bm{x}_0)$ for a new input sample $\bm{x}_0$. Each input sample $\bm{x}_0$ represents a possible set of operating conditions, such as specific values of load power and wind speed.
Fig. \ref{fig:Kriging_explain} illustrates the Kriging method and the main steps discussed in this subsection. The method produces two key results: the Kriging mean in \eqref{eq:K_mean}, which provides the predicted PTSM, and the Kriging variance in \eqref{eq:K_var}, which quantifies the prediction credibility.

\subsubsection{Kriging Prediction of the PTSM Response:}\label{subsubsec:Kriging_mean_var}

Let $\mathcal{D}_{\bm{X}}$ denote the domain of input random variables $\bm{X}$. Kriging, or more specifically, the ordinary Kriging 
assumes that the PTSM response $\mathcal{M}(\bm{x})$ in \eqref{eq:PCCT} is a realization of a Gaussian process (GP) indexed by the uncertain inputs $\bm{x}\in \mathcal{D}_{\bm{X}}\subset\mathbb{R}^M$ \cite{bastos2009diagnostics}:
\begin{equation}
\label{eq:GP_model}
    \mathcal{M}^K(\bm{x}|\beta, \bm{\theta}, \sigma^2_z)\sim \mathrm{GP}(\beta,  \sigma_z^2 R(\bm{x}, \bm{x}^{\prime}| \bm{\theta}))
\end{equation}
where ${\beta}$ denotes the unknown constant mean of the Gaussian process $\mathcal{M}^K(\bm{x})$; 
 $\sigma_{z}^2$ denotes the unknown process variance; $R(\bm{x}, \bm{x}^{\prime}| \bm{\theta})$ is the correlation function between $\bm{x}, \bm{x}^{\prime}\in\mathcal{D}_{\bm{X}}$, with $\bm{\theta}$ denoting its unknown hyperparameters.
In this work, we adopt the ellipsoidal, anisotropic Matérn-5/2 correlation function due to its robustness in modelling real-world physical systems \cite{williams2006gaussian}. 
The estimation of $\beta$ is embedded in the Kriging predictor derivation and is not shown in Fig.~\ref{fig:Kriging_explain}.

With the general Kriging model~\eqref{eq:GP_model} formulated, we next incorporate the training dataset $(\mathcal{X}, \mathcal{T})$ to enable predictive inference of the PTSM response $\mathcal{M}(\bm{x}_0)$, given parameters $\eta=(\beta, \bm{\theta}, \sigma^2_z)$.
Let $\bm{T}_K=[\mathcal{M}^K(\bm{x}^{(1)}|\eta),..., \mathcal{M}^K(\bm{x}^{(N)}|\eta)]^T$ denote the evaluations of $\mathcal{M}^K(\bm{x}|\eta)$ at the training samples $\mathcal{X}=[\bm{x}^{(1)}, \dots, \bm{x}^{(N)}]^T$. Let $\mathcal{M}^{K}(\bm{x}_0|\eta)$ represent the Gaussian process-based prediction of $\mathcal{M}(\bm{x}_0)$. Due to the Gaussian assumption, $\mathcal{M}^{K}(\bm{x}_0|\eta)$ and $\bm{T}_K$ follow a joint Gaussian distribution \cite{Dubourg2011}:
\begin{equation}
    \begin{bmatrix}
\mathcal{M}^K(\bm{x}_0|\eta) \\
\bm{T}_K \\
\end{bmatrix}
\sim
\mathcal{N}_{N+1} \left(
\begin{bmatrix}
\beta \\
\bm{1}\beta 
\end{bmatrix},
\sigma^2_z\begin{bmatrix}
1 & \bm{r}^T(\bm{x}_0)  \\
\bm{r}(\bm{x}_0) & \bm{R}  \\
\end{bmatrix}
\right)
\label{eq:Kriging_basic_eq}
\end{equation}
where the $N\times1$ vector $\bm{r}(\bm{x}_0)$ denotes  the correlations between $\bm{x}_0$ and $\mathcal{X}$, with entries $r_n = R(\bm{x}_0, \bm{x}^{(n)}| \bm{\theta}), n=1,\dots, N$. The $N\times N$ matrix $\bm{R}$ is the matrix of pairwise correlations between the samples in $\mathcal{X}$. $\bm{1}$ is a $N\times1$ vector of all ones.
Assuming $(\bm{\theta}, \sigma^2_z)$ are known, the predictive distribution of $\mathcal{M}^K(\bm{x}_0)$ conditioned on the training data $(\mathcal{X}, \mathcal{T})$ is given by (see Eqs.~(4)-(12) in \cite{bastos2009diagnostics} for details):
\small
\begin{equation}
    \mathcal{M}^K(\bm{x}_0|\mathcal{X}, \bm{T}_K=\mathcal{T}, \bm{\theta}, \sigma_z^2)\sim\mathcal{N}(\mu_{\hat{T}}(\bm{x}_0|\bm{\theta}), \sigma_{\hat{T}}^2(\bm{x}_0| \bm{\theta}, \sigma_z^2))
    \label{eq:Kriging_predictor}
\end{equation}
\normalsize
This distribution is referred to as
the Kriging predictor \cite{Dubourg2011}
for the true PTSM response $\mathcal{M}(\bm{x}_0)$.
Specifically, the Kriging predictor is a Gaussian random variable
with mean $\mu_{\hat{T}}(\bm{x}_0|\bm{\theta})$ and variance $\sigma_{\hat{T}}^2(\bm{x}_0| \bm{\theta}, \sigma_z^2)$, whose closed-form expressions are provided below~\cite{Dubourg2011}:
\begin{eqnarray}
 &\mu_{\hat{T}}(\bm{x}_0|\bm{\theta})=\hat{\beta}+\bm{r}^T(\bm{x}_0)\bm{R}^{-1}(\mathcal{T}-\bm{1}\hat{\beta})
    \label{eq:K_mean}\\
 &   \sigma_{\hat{T}}^2(\bm{x}_0| \bm{\theta}, \sigma_z^2)=\sigma_z^2[1-\bm{r}^T\bm{R}^{-1}\bm{r}+u^2/Q]  
    \label{eq:K_var}\\
  &  \hat{\beta} =(\bm{1}^T\bm{R}^{-1}\bm{1})^{-1}\bm{1}^T\bm{R}^{-1}\mathcal{T} \label{eq:Beta}\color{black}
\end{eqnarray}
where $u=1-\bm{1}^T\bm{R}^{-1}\bm{r}$ and $Q=\bm{1}^T\bm{R}^{-1}\bm{1}$. 
While the above derivation assumes $(\bm{\theta}, \sigma_{z}^2)$ to be known, these parameters must be estimated from data before computing the Kriging mean \eqref{eq:K_mean} and variance \eqref{eq:K_var}.

\subsubsection{Solving the Parameters:}
\label{subsubsec:K_para_est}
First, the correlation hyperparameter $\bm{\theta}$ is  estimated by solving the following leave-one-out cross-validation (LOO-CV) optimization problem \cite{bachoc2013cross}:
\begin{equation}  \hat{\bm{\theta}}=\underset{\bm{\theta} \in \mathbb{R}^M}{\arg \min}\sum_{n=1}^{N}\left(\mu^{-n}_{\hat{T}}(\bm{x}^{(n)}, \bm{\theta})- \mathcal{M}(\bm{x}^{(n)})\right)^2 \label{eq:LOO-XV}
\end{equation}
where $\mu^{-n}_{\hat{T}}(\bm{x}^{(n)}, \bm{\theta})=\mu_{\hat{T}}(\bm{x}^{(n)}, \bm{\theta}| \mathcal{X}_{-n}, \mathcal{T}_{-n})$ denotes the Kriging mean \eqref{eq:K_mean} calculated using the reduced dataset $(\mathcal{X}_{-n}, \mathcal{T}_{-n})$, which is derived from $(\mathcal{X}, \mathcal{T})$ by excluding the $n$-th sample $(\bm{x}^{(n)}, \mathcal{M}(\bm{x}^{(n)}))$.
Owing to its efficiency in handling complex nonlinear optimization tasks,
a hybrid genetic algorithm \cite{Dubourg2011} is employed to solve \eqref{eq:LOO-XV}. 

Given the estimated hyperparameters $\hat{\bm{\theta}}$, the process variance $\sigma_z^2$
is subsequently calculated as \cite{bachoc2013cross}:
\begin{equation} \hat{\sigma}_z^2=\hat{\sigma}_z^2(\hat{\bm{\theta}})=\frac{1}{N}\sum_{n=1}^{N}\frac{\left(\mu^{-n}_{\hat{T}}(\bm{x}^{(n)}, \hat{\bm{\theta}})- \mathcal{M}(\bm{x}^{(n)})\right)^2}{c^{-n}_{\hat{T}}(\bm{x}^{(n)}, \hat{\bm{\theta}})}
  \label{eq:estimated_var}
\end{equation}
where $c^{-n}_{\hat{T}}(\bm{x}^{(n)}, \hat{\bm{\theta}})=[1-\bm{r}^T\bm{R}^{-1}\bm{r}+u^2/Q]$ denotes the normalized Kriging variance in \eqref{eq:K_var}, evaluated at $\bm{x}^{(n)}\in\mathcal{X}$ based on the reduced dataset 
$(\mathcal{X}_{-n}, \mathcal{T}_{-n})$
and the estimated hyperparameters $\hat{\bm{\theta}}$.

Given the estimated parameters ($\hat{\bm{\theta}}, \hat{\sigma}_z^2)$, the Kriging predictor \eqref{eq:Kriging_predictor} can be constructed. This involves computing the Kriging mean  $\mu_{\hat{T}}(\bm{x}_0|\hat{\bm{\theta}})$ and the Kriging variance $\sigma_{\hat{T}}^2(\bm{x}_0|\hat{\bm{\theta}}, \hat{\sigma}_z^2)$ by \eqref{eq:K_mean} and \eqref{eq:K_var}.  
The Kriging mean then serves as a surrogate model to approximate the true response $\mathcal{M}(\bm{x}_0)$: 
\begin{equation}
    \mathcal{M}(\bm{x}_0)\approx \mu_{\hat{T}}(\bm{x}_0|\hat{\bm{\theta}})
    \label{eq:kriging_surrogate}
\end{equation}
The corresponding Kriging variance $\sigma_{\hat{T}}^2(\bm{x}_0|\hat{\bm{\theta}}, \hat{\sigma}_z^2)$ serves as an indicator of prediction credibility. 
\color{black}For example, a large $\sigma_{\hat{T}}^2(\bm{x}_0)$ suggests that $\bm{x}_0$ lies far from the training samples $\mathcal{X}$, implying reduced credibility of the prediction $\mu_{\hat{T}}(\bm{x}_0|\hat{\bm{\theta}})$. 
Additionally, the Kriging predictor \eqref{eq:Kriging_predictor} interpolates the training dataset, i.e., $\mu_{\hat{T}}(\bm{x}_0) = \mathcal{M}(\bm{x}_0)$ and $\sigma_{\hat{T}}^2(\bm{x}_0)=0$ for all $\bm{x}_0 \in \mathcal{X}$.





\subsection{Surrogate Model-Based Reliability Analysis}
\label{subsec:Kriging_based_RA}
The well-trained Kriging mean \eqref{eq:K_mean} serves as a surrogate model for conducting rapid reliability analysis. Let $\mathcal{X}_V$ denote a high-volume sample selection pool (e.g., with $N_V=10^5$ samples) obtained from statistical sampling. 
Then, at the $l$-th iteration of the active learning, the surrogate model-based Monte Carlo estimate of the instability probability is given by: 
\begin{equation}
\hat{P}_f^{(l)} = \frac{1}{N_V}\sum_{\bm{x}\in\mathcal{X}_V}I_{m<0}(\mu_l(\bm{x}))
    \label{eq:Pf_predict_general}
\end{equation}
where $\mu_l, l\in\mathbb{N}^+$ denotes the Kriging mean trained at the $l$-th iteration; $I_{m<0}(m)$ is an indicator function (i.e., $I_{m<0}(m)=1$ when $m<0$. Otherwise, $I_{m<0}(m)=0$). 
\color{black}
In other words, the estimated instability probability in \eqref{eq:Pf_predict_general} is calculated from the predicted TSM values over the sample pool $\mathcal{X}_V$ as the proportion of samples satisfying $\mu_l(\bm{x})<0$, rather than being directly predicted by the surrogate model as a probability output.
\color{black}

\subsection{Stopping Criterion} 
\label{subssec:sc}

While the active learning process enriches the training dataset in each iteration to improve the prediction accuracy of $P_f$, a stopping criterion is necessary to avoid inefficiencies from excessive data enrichment. 
To this end, we adopt a convergence criterion based on the stability of $P_f$ estimates:
\begin{equation}
    \frac{\max\{\hat{P}_f^{(j)}\}-\min\{\hat{P}_f^{(j)}\}}{\min\{\hat{P}_f^{(j)}\}} \leq \epsilon_s,\ j=l-l_{ck}+1,\dots, l 
    \label{eq:StopCrit}
\end{equation}
Here, $\hat{P}_f^{(j)}$ denotes the predicted instability probability at iteration $j$, 
$\epsilon_s$  is the error tolerance, and $l_{ck}$ is the size of the checking window. 
For example, with $l=10$ and $l_{ck}=3$, the active learning will stop at the $10$th iteration if the recent estimates $\{\hat{P}_f^{(j)}\}=\{\hat{P}_f^{(8)}, \hat{P}_f^{(9)}, \hat{P}_f^{(10)}\}$ satisfy \eqref{eq:StopCrit}.

\subsection{Learning Function}
\label{subsec:LF}
If the stopping criterion is not satisfied, a learning function is then used to select the most informative samples from the selection pool $\mathcal{X}_{V}$ to enrich the current training dataset.
The surrogate model retrained on the enriched dataset is expected to yield a more accurate prediction of instability probability.

One widely used learning function for the Kriging model is the U-learning function \cite{echard2011ak}. At the $l$-th iteration, for any $x_0 \in \mathcal{X}_V$, the U-function is defined as:
\begin{equation}
     U_l(\bm{x}_0)=\frac{|\mu_l(\bm{x}_0)|}{\sigma_l(\bm{x}_0)}
    \label{eq:U_function}
\end{equation}
\noindent where $\sigma_l^2$ is the Kriging variance \eqref{eq:K_var} at iteration $l$, and $\sigma_l$ denotes its positive square root. The next sample to be enriched $\bm{x}_{e}\in\mathcal{X}_{V}$ is the one that minimizes the U-function:
\begin{equation}
    \bm{x}_{e}= \argmin_{\bm{x}_0\in\mathcal{X}_{V}} U_l(\bm{x}_0)
    \label{eq:min_U}
\end{equation}
\eqref{eq:min_U} implies that the selected sample $\bm{x}_{e}$ should correspond to a predicted stability margin that is near the transient stability boundary  (i.e., $\mathcal{M}(\bm{x}_{e})\approx\mu_{l}(\bm{x}_{e})\to 0$), exhibit low predictive credibility (i.e., large $\sigma_{l}(\bm{x}_{e})$ value), or satisfy both conditions. 
\color{black}
Most of these newly selected samples are expected to concentrate near the stability boundary, where accurate surrogate prediction is most critical for distinguishing rare unstable samples from stable ones. As a result, the surrogate model $\mathcal{M}_s$ trained on the enriched dataset can achieve improved local accuracy near the stability boundary, thereby enabling more accurate identification of rare unstable regions and estimation of the corresponding probability of instability.
\color{black}

Compared with the entropy-based learning function in \cite{zhang2021power}, the U-function remains effective even when no unstable samples are present in the training dataset $(\mathcal{X}, \mathcal{T})$. This is because it exploits continuous predictive information from the CCT-based stability margins \eqref{eq:PCCT}, rather than relying on discrete class labels.

In this work, multiple samples are enriched at each iteration to improve the efficiency of active learning. Specifically, at iteration $l$, a set $\mathcal{X}^{(l)}_e$ of $n_e$ samples selected by \eqref{eq:min_U} is added to the current training dataset.

\section{The active learning framework for transient instability probability assessment}
\label{sec:algorithm}

\begin{figure}[!h]
\centering
\includegraphics[width=1\linewidth]{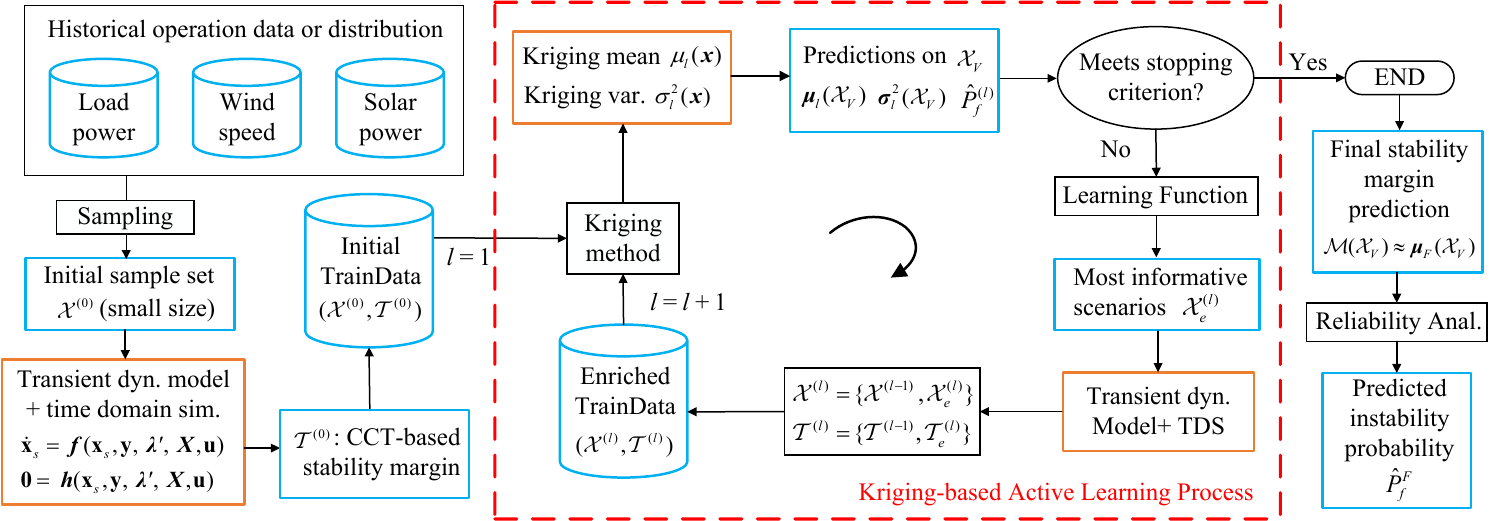}
\caption{
The active learning framework for transient instability probability assessment. Key computational data are outlined in blue, while the computational models are highlighted in orange. \textcolor{black}{The probabilistic nature of this framework lies in the statistical sampling of input scenarios from the uncertainty space and the estimation of instability probability over  $\mathcal{X}_V$.}
}
\label{fig:algorithmn}
\end{figure}

Fig. \ref{fig:algorithmn} provides an overview of the proposed active learning (AL) framework, and Algorithm \ref{alg:general_framework}  details its implementation. 
\color{black}
The objective of this Kriging-based AL framework is to enable efficient transient instability probability assessment under a limited TDS budget by accurately identifying rare instability regions in the input uncertainty space.
\color{black}
Several remarks on surrogate model construction, computational scalability, and algorithm parameters are presented below. 

\color{black}{
\begingroup
\renewcommand{\baselinestretch}{1.2}

\begin{algorithm}

   \caption{\textcolor{black}{The active learning framework for 
   transient instability probability assessment}}

    \begin{algorithmic}[1] \small

   \State \textbf{Input Data \& Model:} Initial training sample set $\mathcal{X}^{(0)}$; Sample selection pool $\mathcal{X}_V$; 
   System transient dynamic model \eqref{eq:Stochastic-PSDAE}.  
   \label{alg_step:input_data}
   \State \textbf{Input Iteration Parameters:} Size of enriched sample set $\mathcal{X}_e^{(l)}$ per iteration, denoted by $n_e$; Iteration limit $l_{max}$; Stopping criterion parameters $(\epsilon_s, l_{ck})$.\label{alg_step:input_para}
   \State Calculate $\mathcal{T}^{(0)}$, the  transient stability margin (TSM) of $\mathcal{X}^{(0)}$, by time-domain simulation (TDS) of \eqref{eq:Stochastic-PSDAE}.\label{alg_step:TDS_initialED}
   \State Assemble the initial training dataset $(\mathcal{X}^{(0)}, \mathcal{T}^{(0)})$. Initialize the iteration index $l=1$; \label{alg_step:Form_initialED}
    \For{$ l= 1, \dots, l_{max}$} \label{alg_step:AL_start}
        \State  Build the Kriging Model based on $(\mathcal{X}^{(l-1)}, \mathcal{T}^{(l-1)})$:
        \label{alg_step:Kriging_training}
   \begin{enumerate}[leftmargin=1cm, label=\alph*)]
        \item  Solve the parameters $(\hat{\bm{\theta}}, \hat{\sigma_z^2})$ by  \eqref{eq:LOO-XV} and \eqref{eq:estimated_var};
        \item  Substitute $(\hat{\bm{\theta}}, \hat{\sigma_z^2}, \mathcal{X}^{(l-1)}, \mathcal{T}^{(l-1)})$ into \eqref{eq:K_mean} and \eqref{eq:K_var}, to obtain the Kriging mean $\mu_l(\bm{x})$ and variance $\sigma^2_l(\bm{x})$;
   \end{enumerate} \label{alg_step:build_K}
   
    \State Predict the TSM of $\mathcal{X}_V$ by the Kriging model $\mu_l(\bm{x})$ in \eqref{eq:K_mean}. The predicted TSM values of  $\mathcal{X}_V$ are denoted as $\mu_l(\mathcal{X}_V)$;\label{alg_step:X_V_Kriging_mean} 
    
    \State  Estimate the instability probability $\hat{P}_f^{(l)}$ by \eqref{eq:Pf_predict_general} based on $\mu_l(\mathcal{X}_V)$;

    \State Compute $\sigma^2_l(\mathcal{X}_V)$, the Kriging variance of $\mathcal{X}_V$, using~\eqref{eq:K_var};
    \label{alg_step:X_V_Kriging_var} 

    \If{$\hat{P}_f^{(l)}$ meets \eqref{eq:StopCrit}} \label{alg_step:stopcrit} 
    
    \State 
    Go to \textbf{Step \ref{alg_step:Gen_results}};
    \EndIf 

    \State Calculate \eqref{eq:U_function} based on $\mu_{l}(\mathcal{X}_V)$ and $\sigma^2_{l}(\mathcal{X}_V)$;
    \State  Identify the set $\mathcal{X}_e^{(l)}$ of size $n_e$ for enrichment by \eqref{eq:min_U}; \label{alg_step:LF_enrichment}
    \State Solve the TSM $\mathcal{T}_e^{(l)}$ of  $\mathcal{X}_e^{(l)}$ by TDS of \eqref{eq:Stochastic-PSDAE}; \label{alg_step:tds}
    \State  $\mathcal{X}^{(l)}= \{\mathcal{X}^{(l-1)}, \mathcal{X}_e^{(l)}\}$ $ , \mathcal{T}^{(l)}=\{\mathcal{T}^{(l-1)}, \mathcal{T}_e^{(l)}\}$; 
    \EndFor  \label{alg_step:AL_end}
     
   \State \label{alg_step:Gen_results} Output the final TSM predictions $\mu_{F}(\mathcal{X}_V)=\mu_{l}(\mathcal{X}_V)$, and the instability probability estimate $\hat{P}_f^F=\hat{P}_f^{(l)}$.

   \end{algorithmic}
   
    \label{alg:general_framework}
\end{algorithm}
\endgroup
}

\color{black}
\noindent \textbf{Remarks on the training of Kriging model (Step \ref{alg_step:Kriging_training}):} The Kriging model is iteratively retrained to obtain a more accurate approximation of the boundary of the rare instability region in the uncertainty space. Accordingly, the input features of each sample consist of the specific values of all modeled uncertainty sources, such as wind speed, load level, and other uncertain operating-condition variables considered in the case study. Each sample is distinguished by its own combination of these input values. The training dataset at iteration $l$, denoted by $(\mathcal{X}^{(l)}, \mathcal{T}^{(l)})$, is constructed by enriching the dataset from the previous iteration, $(\mathcal{X}^{(l-1)}, \mathcal{T}^{(l-1)})$, with newly selected samples near the stability boundary from the enrichment pool $\mathcal{X}_V$. Both the initial sample set $\mathcal{X}^{(0)}$ and the enrichment pool $\mathcal{X}_V$ are generated through statistical sampling from the probability distributions of the uncertain inputs. For each selected input sample, the corresponding output target is the transient stability margin, which is calculated via TDS.
\color{black}

\noindent \textbf{Remarks on the size of $\mathcal{X}_V$ and $\mathcal{X}^{(0)}$ (Step \ref{alg_step:input_data}):}
The initial sample set $\mathcal{X}^{(0)}$ should be kept small (e.g., 50 samples), since Kriging only requires a small training dataset $(\mathcal{X}^{(0)}, \mathcal{T}^{(0)})$ to build an initial surrogate model for subsequent active learning.
\color{black}
The sample selection pool $\mathcal{X}_V$ should be sufficiently large to contain an adequate number of unstable samples. Therefore, its size $N_V$ should be chosen according to the level of the instability probability $P_f$. For example, when $P_f \approx 1\%$, $N_V=10^5$ is sufficient, as it yields around $10^3$ unstable samples. The impact of $N_V$ will be further discussed in Part \ref{subsubsec:PoolSizeTest} of Section \ref{subsec:ParameterAnalysis}. Despite its large size, constructing $\mathcal{X}_V$ requires little computational effort, since it only involves sampling the input space without performing time-domain simulations.
\color{black}
 Step \ref{alg_step:X_V_Kriging_mean} and \ref{alg_step:X_V_Kriging_var} based on $\mathcal{X}_V$ take negligible time, as the Kriging mean \eqref{eq:K_mean} and variance \eqref{eq:K_var} are cheap to evaluate. 

\noindent\noindent\textbf{Remarks on the Iteration Computational Cost and Scalability:}
The computational cost of the proposed Kriging-based AL framework (AL-Kriging) is primarily twofold: (i) the cost associated with TDS of the enriched samples (Step \ref{alg_step:tds}); (ii) the cost associated with the iterative construction of Kriging models (Step \ref{alg_step:build_K}).
The TDS computational cost can be effectively limited by the maximum number of enriched samples, given by $n_e(l_{max} - 1)$.
The Kriging training complexity scales cubically with the number of training samples $N$, i.e., $\mathcal{O}(N^3)$ \cite{liu2020gaussian}. Further efficiency improvement could be achieved using advanced methods such as sparse Kriging 
\cite{liu2020gaussian}. 

That being said, the training cost of AL-Kriging is not sensitive to the system size or the number of random inputs. 
Instead, it depends on the complexity of the instability domain
associated with specific contingencies and systems, which is nevertheless hard to quantify. As the instability domain becomes more complex, more training samples may be required, leading to increased total computational time.
As will be shown in the case studies (59-bus and 240-bus systems), the training cost of AL-Kriging remains similar across system sizes and input dimensionalities. 
However, it varies among contingencies due to differences in instability domain complexity.  



\noindent\textbf{Remarks on the Iteration Parameters (Step \ref{alg_step:input_para}):} The parameters in Table \ref{tab:iter_para} are chosen to balance prediction accuracy and computational cost during iterative Kriging retraining.
First, the Kriging model is retrained every $n_e = 10$ enriched samples. Reducing $n_e$ (e.g., to 5) 
yields only a marginal improvement 
while approximately doubling the retraining cost. 
Second, $l_{max}$ is set to 40 based on our observation that $P_f$ prediction accuracy often saturates 
when $n_e(l_{max} - 1) \geq 400$. Finally, the stopping criterion in Step \ref{alg_step:stopcrit} prevents excessive TDS runs and retraining. 
However, early iterations often yield noisy estimates of $P_f$. To avoid the premature termination of the active learning, we set $\epsilon_s = 0.02$ and $l_{ck} = 5$. Empirically, when $n_e=10$, setting $l_{ck} \geq 3$ and $\epsilon_s \leq 0.05$ is effective in preventing early termination. 
It is worth noting that the same parameters are used across case studies to demonstrate the robustness of the method to parameter choices. 
\color{black}
The impact of these parameters on the efficiency of AL-Kriging will be discussed in Section \ref{subsec:ParameterAnalysis}.
\color{black}
\vspace{-0.3cm}
\begin{table}[h]     
\renewcommand{\arraystretch}{1.1}
    \caption{Configurations of the active learning iteration parameters}
    \label{tab:iter_para}
    \centering
\resizebox{\linewidth}{!}{    
    \begin{tabular}{ccc}
    \hline
          Parameters         & Description      &   Value  \\ \hline
      $n_e$ & Size of the $l$-th sample enrichment $\mathcal{X}_e^{(l)}$ (Step \ref{alg_step:LF_enrichment})    & $10$ \\ 
      
      $\epsilon_s$  & Stopping criterion \eqref{eq:StopCrit} error tolerance (Step \ref{alg_step:stopcrit})             & $0.02$ \\ 
      
      $l_{ck}$  &  Stopping criterion \eqref{eq:StopCrit} checking range (Step \ref{alg_step:stopcrit})              & $5$\\ 
      $l_{max}$  &  Maximum number of iterations (Step \ref{alg_step:AL_start})        & $40$  \\ 
     $n_e(l_{max}-1)$  & Maximum number of enriched samples & 390 \\ \hline
    \end{tabular}}
\end{table}

\color{black}
\noindent\textbf{Remarks on the final TSM predictions $\mu_{F}(\mathcal{X}_V)$ (Step \ref{alg_step:Gen_results}):} The final Kriging model $\mu_{F}(\bm{x})$ is trained on a dataset that has been iteratively enriched with samples near the stability boundary under the guidance of the learning function. Therefore, the final TSM predictions $\mu_{F}(\mathcal{X}_V)$ are expected to only have high accuracy near the stability boundary, which is the most critical region for distinguishing rare unstable samples from a large candidate pool $\mathcal{X}_V$ without incurring excessive TDS runs. Accordingly, these final TSM predictions can be reliably used for stability classification, i.e., to identify stable and unstable samples according to the sign of the predicted TSM, but they should not be interpreted as globally accurate pointwise TSM regression results over the entire input space. 

These results reflect the mechanism of the proposed AL-Kriging framework. Through the interaction between the regression surrogate and the learning function, the limited TDS budget is preferentially allocated to informative samples near the stability boundary, thereby supporting efficient identification of rare instability regions in the input uncertainty space.

\color{black}

\section{Case Studies}  
\label{sec:case_study}
The proposed Kriging-based active learning framework (AL-Kriging) is tested in two case studies: the IEEE 59-bus test system \cite{powertech_tsat_manual} and the WECC 240-bus test system \cite{240BusRef}.  
The uncertainties from load power, solar power, and wind speed are considered in our study. Specifically, the relationship between wind speed and the power output of a wind farm is described as follows \cite{lu2024advanced}:
\small
\begin{equation}
    P_w(v) = 
    \begin{cases} 
    P_r \left(\frac{v^3 - v_{in}^3}{v_r^3 - v_{in}^3}\right), & v_{in} \leq v \leq v_r \\
    P_r, & v_r \leq v \leq v_{out} \\
    0, & \text{otherwise}
    \end{cases}
    \label{eq:wind_power_curve}
\end{equation}
\normalsize
where $P_r$, $v_{in}$, $v_{out}$, and $v_r$  denote the rated power, cut-in wind speed, cut-out wind speed, and rated wind speed of the wind turbine generators, respectively. The values of $v_{in}$, $v_{out}$, and $v_r$ are set to 3 m/s, 25 m/s, and 12 m/s \cite{WTG_Parameter_Ref}. 

\color{black}
Given that the proposed AL-Kriging method aims to accurately identify rare unstable samples under a limited TDS budget, its performance is compared primarily with methods requiring a similar number of TDS runs. These baseline methods include the random forest-based active learning method (AL-RF) \cite{zhang2021power} and three surrogate models without active learning (non-AL methods), namely standard Kriging \cite{xu2020probabilistic}, data-driven polynomial chaos expansion (DDPCE) \cite{wang2020data}, and the multilayer perceptron neural network (MLP) \cite{tan2023debiased}. It should be noted that existing deep learning-based methods \cite{tan2019deep, zhan2023hybrid, liu2026two} are generally developed for a different problem setup, which assumes the availability of a large TDS-labeled training dataset and often relies on additional data-imbalance correction techniques. Therefore, among the deep learning-based methods, only the MLP is included in this work as a simple reference baseline.
\color{black}
The AL-RF uses $100$ decision trees and entropy-based learning functions \cite{zhang2021power}. Its number of iterations and enriched samples are matched to those of AL-Kriging for a fair comparison. For the non-AL methods, standard Kriging models are built under the same settings as AL-Kriging; DDPCE models are constructed sparsely and adaptively \cite{liu2022sparse};
MLP hyperparameters, including the number of hidden layers and units, are selected via grid search over candidate values.
\color{black}
All models are trained directly on raw data without requiring any prior knowledge of the input distributions.

The TDS-MCS results based on  $10^5$ TDS runs are treated as the ground-truth benchmark in the subsequent comparisons. This sample size is chosen such that the coefficient of variation of the estimated instability probability $\hat{P}_f$ remains 
 below $5\%$ when $P_f\approx1\%$ 
(see (2.96)-(2.98) in \cite{au2014SusBook})\color{black}.
 
The baseline time-domain simulation of \eqref{eq:Stochastic-PSDAE} is conducted using TSAT/DSATools with a simulation duration of $12$s. 
The Kriging 
models are implemented via the UQLab toolbox \cite{marelli2014uqlab} in MATLAB R2023b, while the random forest and MLP models are implemented using the scikit-learn library \cite{scikit-learn}.
\color{black}
All computations are performed on an Intel Xeon Gold 6126 2.60GHz CPU.

\subsection{Simulation Results on the IEEE 59-Bus System} \label{subsec:59bus_study}

\begin{figure}[!h]
    \centering    \includegraphics[width=0.7\linewidth]{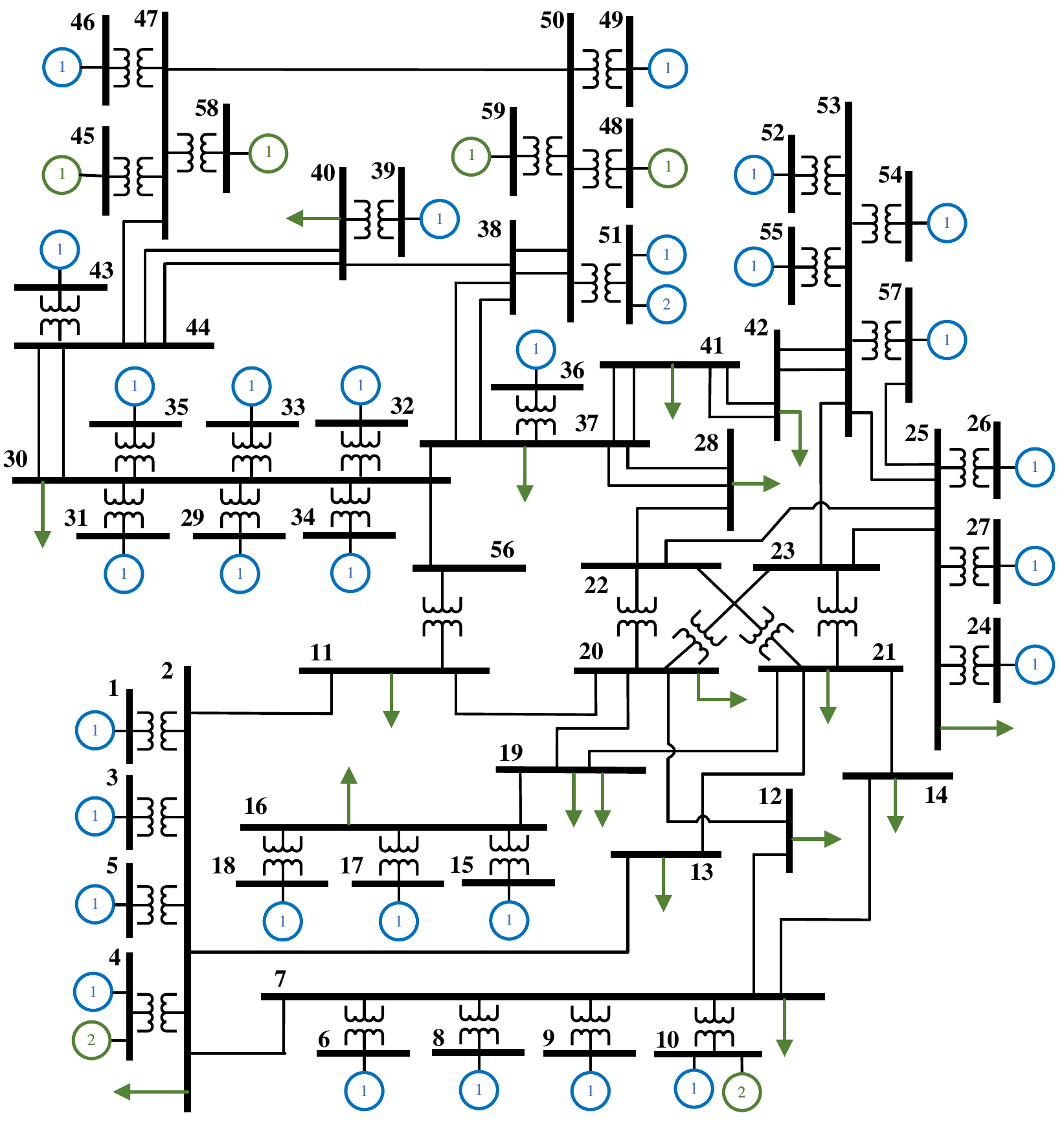}
    \caption{Single line diagram of the modified IEEE 59-bus system with 17 uncertain loads and 6 wind farms. The green arrows indicate the random loads, and the green circles denote the wind farms. }
    \label{fig:59bus}
    \vspace{-0.3cm}
\end{figure}

In this section, we validate and compare the proposed AL-Kriging method against the five other methods 
on the modified IEEE 59-bus test system \cite{powertech_tsat_manual}, as depicted in Fig. \ref{fig:59bus}.

In our study, all 17 loads are random. The active power of the loads is assumed to follow a Gaussian distribution $\mathcal{N}(1, 0.05^2)$ of its base power \cite{xu2020probabilistic},
and the load power factors remain unchanged. The system comprises 6 wind farms, each formed by GE 1.5 MW wind turbine generators \cite{GE1p5MW_manual}, with a total rated power of 51 MW. The wind speed is assumed to follow a Weibull distribution, with a scale parameter $11.2$ and
a shape parameter $2.2$ \cite{krogsaeter2015validation}. The 17 random loads and 6 wind farms are located at Buses (2, 7, 11-14, 16, 19, 20, 21, 25, 28, 30, 37, 40, 41) and Buses (4, 10, 45, 48, 58, 59), respectively. 
The dependencies among the 23 random inputs are modelled by the Gaussian copula \cite{sheng2018applying}. To simplify the dependency structure, the inputs are partitioned into five independent groups. Inputs within the same group are correlated, whereas inputs from different groups are considered independent. Table \ref{tab:59BusCorGroup} lists the groupings and their correlation coefficients
\cite{ye2022stochastic} \cite{yue2020probabilistic}.
\color{black}
These 23 uncertain wind and load power variables are used as the input features of each sample for surrogate model training, while the corresponding output target is the transient stability margin.
\color{black}

Under the specified uncertainty configurations, the performance of the proposed method and other surrogate models was evaluated across the 6 contingencies listed in Table \ref{tab:59BusCtg}. Each contingency involves a three-phase ground fault of duration $t_{\mathrm{fct}}$, cleared by tripping a specific transmission line.

\begin{table}[t]        
    \caption{Summary of the independent groups}
    \label{tab:59BusCorGroup}
    \centering
    \begin{tabular}{ccc}
    \hline
    Group name & Description & Corr. coeff.  \\ \hline
     Load power (Bus 1-21)  & Loads at 138kV  & 0.4 \\ 
      Load power (Bus 22-59)  &  Loads at 230kV & 0.2\\ 
      Wind speed (Bus 45 \& 58) &  Connected to Bus 47 & 0.8  \\ 
      Wind speed (Bus 48 \& 59) & Connected to Bus 50  & 0.8 \\ 
      Wind speed (Bus 4 \& 10) & Connected to 138kV grid  & 0.8 \\ \hline       
    \end{tabular}
    \vspace{-0.3cm}
\end{table}

\begin{table}[t]
\caption{List of the tested contingencies (59-Bus system)}
\label{tab:59BusCtg}
\centering
\begin{tabular}{cccc}
\hline
Case & 3-phase fault loc. &  Line tripped & $t_{\mathrm{fct}}$ (cycle) \\ \hline
C1 & Bus 19 & 19-21 & $8$  \\   
C2 & Bus 19 & 19-20 & $5$  \\   
C3 & Bus 38 & 38-50 & $8$  \\   
C4 & Bus 38 & 38-37 & $8$  \\   
C5 & Bus 30 & 30-44 & $8$  \\   
C6 & Bus 30 & 30-56 & $8$  \\  \hline 
\end{tabular}
{\scriptsize
\begin{tablenotes}
\item{*} For example, contingency C1 means the three-phase ground fault at Bus 19, which is cleared after $8$ cycles by tripping Line 19-21.
\end{tablenotes}
}
\end{table}

\subsubsection{Training 
of the  AL-based Surrogate Model:}
\label{subsubsec:al_para_59}
First, an initial training dataset $(\mathcal{X}^{(0)}, \mathcal{T}^{(0)})$ of size $N=50$ is generated via Latin hypercube sampling and corresponding time-domain simulations (Algorithm \ref{alg:general_framework}, Step \ref{alg_step:TDS_initialED} and \ref{alg_step:Form_initialED}). This dataset often contains no unstable samples when $P_f$ is small (e.g., around $1\%$). Next, Kriging models are iteratively retrained by enriching samples from a selection pool $\mathcal{X}_V$ of size $N_V=10^5$ (Step \ref{alg_step:AL_start}-\ref{alg_step:AL_end}). The algorithm control parameter settings in Table \ref{tab:iter_para} are adopted. The AL-RF method follows a similar procedure but employs an entropy-based learning function. \color{black} Note that constructing the large selection pool $\mathcal{X}_V$ takes little computational effort, as it does not require time-domain simulations.\color{black}


\subsubsection{Training of the Non-AL Models:}
\label{subsubsec:non_al_para_59}
For comparison, a training dataset $(\mathcal{X}_c, \mathcal{T}_c)$ of size $N_c = 500$ is used to train the non-AL models, namely Kriging, DDPCE, and MLP. 
Since deep neural network-based methods tend to rely on larger datasets, the MLP is further evaluated using a training set of size $5000$.


\subsubsection{Performance Comparison on Instability Probability Prediction:} 
\begin{figure}[t]
    \centering    \includegraphics[width=0.8\linewidth]{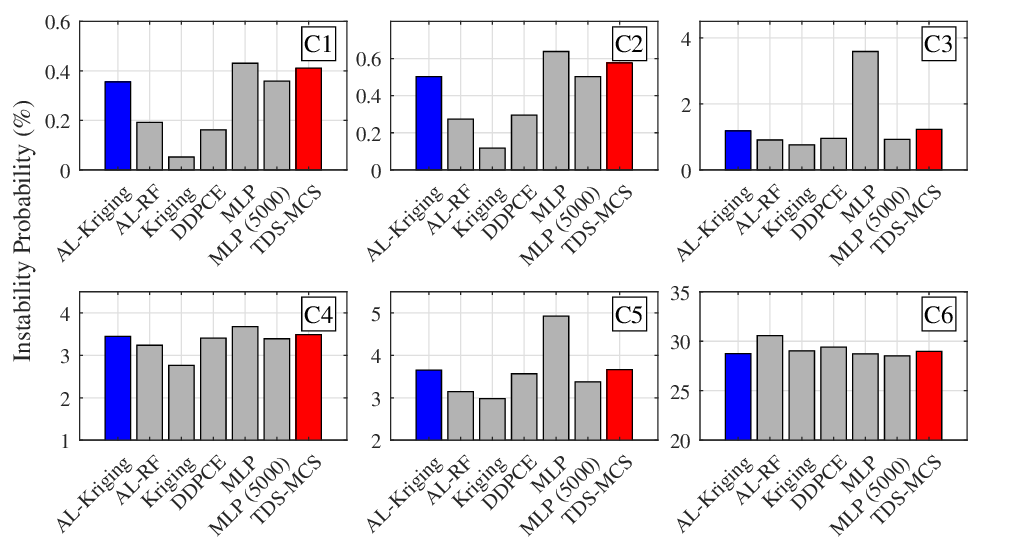}
    \caption{Comparison of the predicted instability probability. 
    MLP(5000) indicates training with 5000 samples; all other non-AL models use 500 samples.
    The benchmark TDS-MCS results are highlighted in red, whereas the results by the proposed AL-Kriging, highlighted in blue,  are closest to those by TDS-MCS. 
    The benchmark instability probabilities for C1-C6 are $0.41\%$, $0.57\%$, $1.23\%$, $3.49\%$, $3.66\%$, and $28.97\%$, respectively.}
    \label{fig:59BUsTIPorb}
\end{figure}

For each contingency in Table \ref{tab:59BusCtg}, Fig. \ref{fig:59BUsTIPorb} presents the instability probabilities predicted by the surrogate models and the benchmark TDS-MCS results.



From Fig. \ref{fig:59BUsTIPorb}, we can make the following observations: First, the tested contingencies can be classified into four levels based on their values of $P_f$. Namely, C1 and C2 ($P_f\approx0.5\%$); C3 ($P_f\approx1\%$); C4 and C5 ($P_f\approx3\%$); C6 ($P_f\approx30\%$). Second, for each contingency, the proposed AL-Kriging consistently achieves decent prediction accuracy of $P_f$, 
especially when $P_f$ is small (e.g., C1-C3). 
While MLP performs comparably for C1–C2, it fails to capture the instability domain accurately, as evidenced by its poor pointwise prediction performance (Table \ref{tab:tpr_fdr_59}; to be discussed).
This limitation of MLP persists even with an order-of-magnitude increase in training size (5000 samples).
The performance advantage of AL-Kriging at small $P_f$ is attributed to the instability-boundary-oriented sample enrichment strategy \eqref{eq:min_U}. Third, although AL-RF also tried fine-tuning the RF model with enriched samples, it cannot accurately predict small $P_f$ values. This is likely due to the entropy-based learning function,  which is ineffective when no unstable samples are present in the initial training dataset $(\mathcal{X}^{(0)}, \mathcal{T}^{(0)})$. Lastly,  all methods perform well when the instability probability is sufficiently large (e.g., C6), which may be attributed to the high proportion of unstable samples in the training dataset, enabling the surrogate models to effectively capture the domain of instability.
\color{black}

\subsubsection{Performance Comparison on Pointwise Rare Instability Events Detection:}

Accurate $P_f$ prediction does not necessarily imply reliable pointwise detection of unstable samples. Therefore, we further compare the pointwise prediction performance of the surrogate models. The TDS-MCS results based on $\mathcal{X}_V$ (sample size $10^5$) are still used as the benchmark. Detection accuracy is quantified by the true positive rate (TPR) and the false discovery rate (FDR) associated with the class confusion matrix \cite{pml1Book} defined in Table \ref{tab:confusion_matrix}. The formulations of TPR and FDR are as follows:
\begin{equation}
\mathrm{TPR} = \frac{N_{\mathrm{TP}}}{N_{\mathrm{TP}}+N_{\mathrm{FN}}} 
\quad
\mathrm{FDR} = \frac{N_{\mathrm{FP}}}{N_{\mathrm{FP}}+N_{\mathrm{TP}}}
\label{eq:TPR_FDR}
\end{equation}
where $N_{\mathrm{TP}}$, $N_{\mathrm{FN}}$, and $N_{\mathrm{FP}}$ denote the number of samples in $\mathcal{X}_V$ identified as True Positive, False Negative, and False Positive (see Table \ref{tab:confusion_matrix}), respectively. 
\color{black}
A higher TPR means that more unstable samples are correctly identified, whereas a lower FDR means that fewer false alarms are produced. Therefore, a desirable surrogate model should achieve a high TPR and a low FDR.
\begin{table}[!]
\caption{Confusion matrix for pointwise detection of unstable samples}
\label{tab:confusion_matrix}
\centering
\color{black}{
\scalebox{1}{
\begin{tabular}{ccc}
\hline
& \makecell[c]{\textbf{Predicted unstable}\\ (by surrogate model)}
& \makecell[c]{\textbf{Predicted stable}\\ (by surrogate model)} \\ \hline
\makecell[c]{\textbf{Actual unstable}\\ (by TDS)}
& True Positive (TP)
& False Negative (FN) \\
\makecell[c]{\textbf{Actual stable}\\ (by TDS)}
& False Positive (FP)
& True Negative (TN) \\ \hline
\end{tabular}
}
}
\begin{tablenotes}
\scriptsize
\item * Unstable samples are treated as the positive class.
\end{tablenotes}
\end{table}
\color{black}

Table \ref{tab:tpr_fdr_59} compares the TPR and FDR of all the methods across the contingencies. 
The results show that the proposed AL-Kriging method consistently identifies most of the unstable samples in the large sample set $\mathcal{X}_V$ (high TPR) while producing only a few false alarms (low FDR), demonstrating its reliable pointwise detection capability, particularly for rare instability events (e.g., C1–C3 with $P_f<2\%$). These identified unstable samples can provide valuable insights into the operating conditions of uncertain loads or RES that may cause rare transient instability.
For example, the unstable samples under C1 are typically associated with high load power at Buses 1–21 and low wind generation at Buses 4 and 10. 
\begin{table}[h!]
\centering
\caption{Comparison of the TPR and FDR \eqref{eq:TPR_FDR} for Contingencies C1--C6 (59-bus system)}
\label{tab:tpr_fdr_59}
\scalebox{0.87}{
\renewcommand{\arraystretch}{1.2}
\begin{tabular}{cccccccc}
\hline
\textbf{Case} & \textbf{Metric} & \textbf{\makecell[c]{AL-\\Kriging}} & \textbf{AL-RF} & \textbf{Kriging} & \textbf{DDPCE} & \textbf{MLP} & \textbf{\makecell[c]{MLP\\(5000)}} \\
\hline
\multirow{2}{*}{C1} & TPR & \textbf{79.3\%} & 42.8\% & 10.2\% & 30.7\% & 59.4\% & 58.6\% \\
                    & FDR & 8.4\% & \textbf{8.3\%} & 19.2\% & 22.2\% & 43.4\% & 32.9\% \\
\multirow{2}{*}{C2} & TPR & \textbf{74.0\%} & 35.5\% & 14.6\% & 44.4\% & 60.1\% & 67.8\% \\
                    & FDR & 15.1\% & 24.9\% & 28.2\% & \textbf{13.2\%} & 45.6\% & 22.3\% \\
\multirow{2}{*}{C3} & TPR & \textbf{90.3\%} & 61.5\% & 53.5\% & 75.7\% & 80.0\% & 72.3\% \\
                    & FDR & 6.2\% & 15.8\% & 11.2\% & \textbf{2.2\%} & 72.2\% & 2.3\% \\
\multirow{2}{*}{C4} & TPR & 91.8\% & 72.1\% & 70.8\% & \textbf{94.5\%} &68.7\% & 92.6\% \\
                    & FDR & 5.0\% & 21.0\% & 9.0\% & \textbf{1.8\%} & 33.3\% & 3.0\% \\
\multirow{2}{*}{C5} & TPR & \textbf{94.7\%} & 69.8\% & 74.7\% & 94.6\% & 78.5\% & 88.8\% \\
                    & FDR & 3.3\% & 17.3\% & 6.9\% & \textbf{1.6\%} & 40.4\% & 2.5\% \\
\multirow{2}{*}{C6} & TPR & 95.7\% & 92.3\% & 94.5\% & 93.8\% & 89.9\% & \textbf{96.0\%} \\
                    & FDR & 3.1\% & 12.1\% & 5.3\% & 7.1\% & 9.0\% & \textbf{2.3\%} \\
\hline

\end{tabular}
}
\vspace{1mm}
{\scriptsize
\begin{tablenotes}
    \item(1) \textbf{MLP(5000)}: Additional test of MLP model trained with 5000 samples due to its higher data demand. All other non-AL models are trained with 500 samples.
    \item(2) \textbf{Higher TPR is preferred}: the surrogate model detects more transient instability events labelled by the benchmark TDS; \textbf{Lower FDR is preferred}: the surrogate model produces fewer false alarms in predicting transient instability events.
\end{tablenotes}
}
\vspace{-0.3cm}
\end{table}

For those contingencies with larger $P_f$ values, some non-AL methods can also perform well and occasionally outperform AL-Kriging 
(e.g., DDPCE on C4). However, AL-RF consistently underperforms AL-Kriging. MLP achieves accuracy comparable to AL-Kriging but requires an order-of-magnitude larger dataset (5000 samples), which is costly in transient stability assessment.
\color{black}
Overall, these results highlight the effectiveness of AL-Kriging in accurately detecting rare instability events with limited training data. 




\subsubsection{Performance Comparison on Computational Cost:}
\begin{figure}[t]
    \centering    \includegraphics[width=0.8\linewidth]{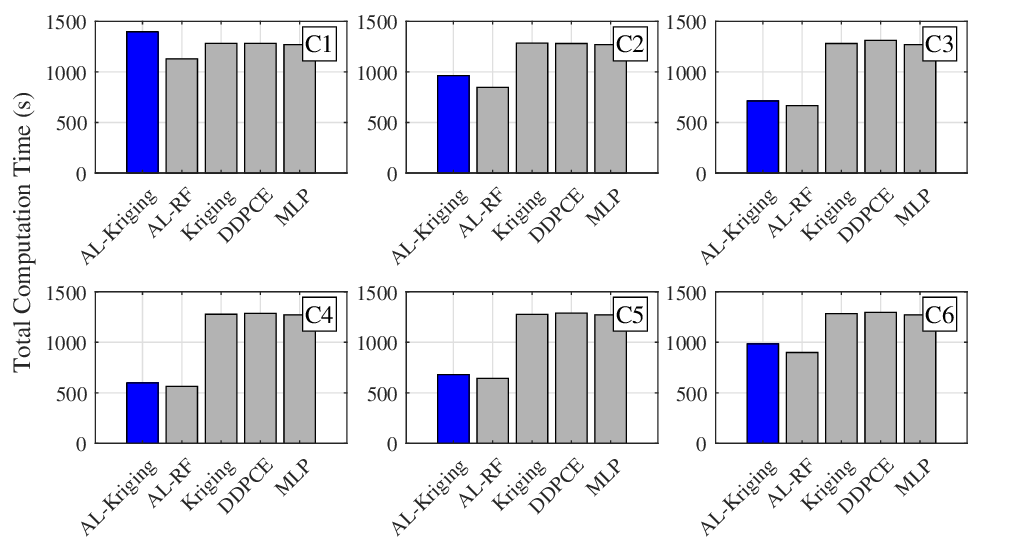}
    \vspace{-0.5cm}
    \caption{The total computation time ($t_{\text{total}}$) of each method  under contingencies listed in Table \ref{tab:59BusCtg}.} 
    \label{fig:59BusTotalTime}
\end{figure}

\begin{table}[!]
\caption{Computation Cost of the AL-Kriging Method (59-Bus System)}
\label{tab:ALK_Time_59Bus}
\centering
\begin{tabular}{ccccccc}
\hline
Case & C1 &  C2 & C3  & C4 & C5 & C6 \\ \hline
$l_{\mathrm{total}}$ & $40$  & $29$  & $22$  & $18$  & $21$  & $31$   \\  
$N_{\mathrm{total}}$ & $440$  &  $330$  &  $260$ &  $220$ & $250$  & $350$   \\ 
$t_{\mathrm{ed}}$(s)  & $1124.9$  & $843.6$  & $664.7$  & $562.4$  &  $639.1$ &  $894.8$  \\  
$t_{\mathrm{sg}}$(s)  &  $273.1$ & $119.7$  &  $48.1$ &  $34.7$ &  $41.3$ & $89.7$ \\  
$t_{\mathrm{total}}$(s)  &  $1397.9$ & $963.3$  &  $712.8$ &  $597.2$ &  $680.5$ & $984.5$ \\  \hline
\end{tabular}
{\scriptsize
\begin{tablenotes}
\item \textbf{Notation:} $l_{\mathrm{total}}$: total number of iterations of the AL process;  $N_{\mathrm{total}}$: total count of time-domain simulation (TDS) executions; $t_{\mathrm{ed}}$: processing time of $N_{\mathrm{total}}$ TDS executions; $t_{\mathrm{sg}}$: time for building the surrogate model; The total time $t_{\text{total}} = t_{\text{sg}} + t_{\text{ed}}$.
\end{tablenotes}
}
\end{table}
Fig.  \ref{fig:59BusTotalTime} shows that AL-RF \cite{zhang2021power} requires the least amount of computational time in all contingencies, but is not accurate as presented in Table \ref{tab:tpr_fdr_59}. 
The AL-Kriging method is more efficient 
than the non-AL methods
\color{black}
in all cases except for C1.
The reason for this exception can be seen in Table \ref{tab:ALK_Time_59Bus}: the AL-Kriging method reaches its iteration limit $l_{max}=40$ for C1, indicating a slow convergence rate of the active learning process. This slow convergence speed could be attributed to 
the complexity of the instability domain $\mathcal{F}$,
\color{black}
as will be discussed in Part \ref{subsubsec:IterLimitTest} of Section \ref{subsec:ParameterAnalysis}.
\color{black}
Table \ref{tab:ALK_Time_59Bus} also shows that the surrogate model construction time $t_{\mathrm{sg}}$ increases with the total number of iterations $l_\mathrm{total}$. This is because the Kriging model is retrained at each step using an expanding dataset  $(\mathcal{X}^{(l)}, \mathcal{T}^{(l)})$, resulting in an increasing training cost per Kriging model. 
Finally, note that for all methods, $t_{\mathrm{ed}}$, the processing time 
of time-domain simulations (TDS) always dominates the total computation time $t_{\mathrm{total}}$, where each TDS execution takes around $2.5$s.

\vspace{-0.2cm}
\subsection{Simulation Results on the WECC 240-Bus System}

The performance of the proposed AL-Kriging is further validated on the modified WECC 240-bus test system, a larger system with detailed dynamic models of conventional generators, wind farms, and solar farms \cite{240BusRef}. 
Specifically, the test system adopts second-generation RES generic models \cite{pourbeik2016generic} for wind and solar farms.
In this case study, the active power outputs of 16 wind farms and 19 solar farms are treated as uncertain. Instead of assuming predefined probability distributions, the random wind speed and solar power are drawn from year-long historical data with a 5-minute resolution \cite{WTG_Parameter_Ref} \cite{NREL_SolarData}. 
\color{black}
The active power outputs of these 16 wind farms and 19 solar farms are used as the input features, and the corresponding output target is the transient stability margin.
\color{black}

Under these uncertainty configurations, we evaluate the performance of AL-Kriging across the contingencies in Table \ref{tab:240BusCtg}. 
To specifically assess AL-Kriging on rare instability events, the fault clearing times $t_{\mathrm{fct}}$ in Table \ref{tab:240BusCtg} are selected so that the resulting instability probabilities fall below 1\%.
\begin{table}[h]
\caption{List of the tested contingencies (240-Bus system)}
\label{tab:240BusCtg}
\centering
\begin{tabular}{cccc}
\hline
Ctg No. & 3-phase fault loc. &  Line tripped & $t_{\mathrm{fct}}$ (cycle) \\ \hline
C1 & Bus 6102 & 6102-6403 & $6$  \\   
C2 & Bus 1401 & 1401-2400 & $8$  \\   
C3 & Bus 3903 & 3903-3923 & $4$  \\   
C4 & Bus 1002 & 1002-1102 & $6$  \\  \hline 
\end{tabular}
{\scriptsize
\begin{tablenotes}
\item{*}  For example, contingency C1 means the three-phase ground fault at Bus 6102, which is cleared after $6$ cycles by tripping Line 6102-6403.
\end{tablenotes}
}
\end{table}

\subsubsection{Training of the AL and the Non-AL Models:} 
To demonstrate the insensitivity of the proposed AL-Kriging method to training parameters and to ensure fair comparison, all model training configurations from the previous case study (Section \ref{subsec:59bus_study}) 
are maintained for both AL and non-AL methods. 
\color{black}



\subsubsection{Comparison of Probability Prediction and Pointwise Instability Event Detection Accuracy:}
\begin{figure}[!]
    \centering    \includegraphics[width=0.8\columnwidth]{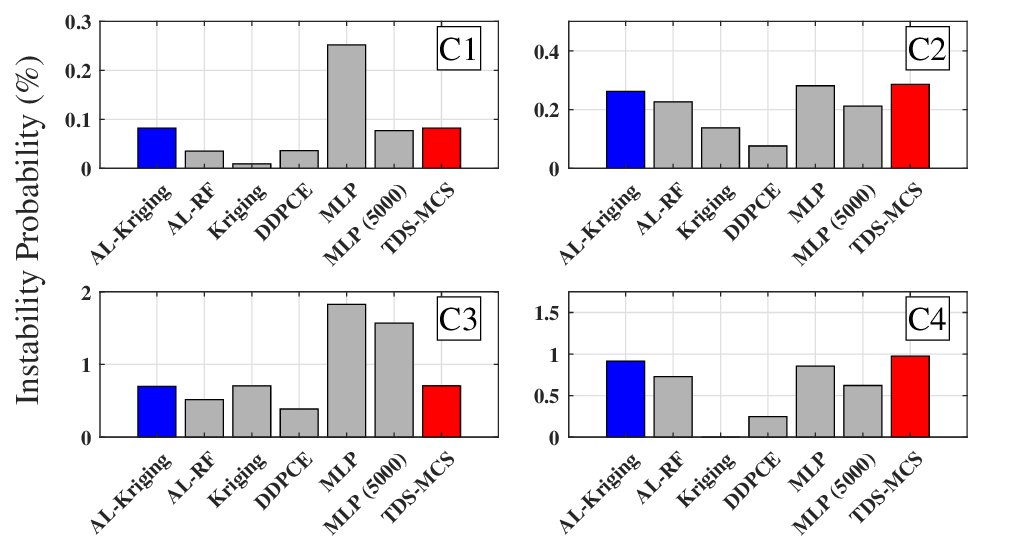}
    \caption{Comparison of the predicted instability probabilities. MLP(5000) indicates training with 5000 samples; all other non-AL models use 500 samples.
    The benchmark TDS-MCS results are highlighted in red. The predictions by the proposed AL-Kriging are highlighted in blue. The benchmark instability probabilities for C1-C4 are $0.08\%$, $0.29\%$, $0.70\%$, and $0.97\%$, respectively. }
    \label{fig:240BusTrFPorb}
\vspace{-0.4cm}
\end{figure}

For all contingencies in Table \ref{tab:240BusCtg}, Fig. \ref{fig:240BusTrFPorb} summarizes the predicted instability probabilities, and Table \ref{tab:tpr_fdr_240} presents the pointwise detection accuracy of all methods.
\color{black}
\begin{table}[!]
\centering
\caption{Comparison of the TPR and FDR \eqref{eq:TPR_FDR} for Contingencies C1--C4 (240-Bus System)}
\label{tab:tpr_fdr_240}
\scalebox{0.87}{
\renewcommand{\arraystretch}{1.2}
\begin{tabular}{cccccccc}
\hline
\textbf{Case} & \textbf{Metric} & \textbf{\makecell[c]{AL-\\Kriging}} & \textbf{AL-RF} & \textbf{Kriging} & \textbf{DDPCE} & \textbf{MLP} & \textbf{\makecell[c]{MLP\\(5000)}} \\
\hline
\multirow{2}{*}{C1} & TPR & \textbf{100.0\%} & 42.7\% & 11.0\% & 43.9\% & 76.8\% & 82.9\% \\
                    & FDR & \textbf{0.0\%} & 0.0\% & 0.0\% & 0.0\% & 75.0\% & 11.7\% \\
\multirow{2}{*}{C2} & TPR & \textbf{91.6\%} & 77.6\% & 41.6\% & 10.5\% & 36.7\% & 66.4\% \\
                    & FDR & \textbf{0.0\%} & 1.8\% & 13.8\% & 60.5\% & 62.2\% & 9.1\% \\
\multirow{2}{*}{C3} & TPR & 98.3\% & 63.2\% & 71.0\% & 31.7\% & \textbf{99.4\%} & 82.0\% \\
                    & FDR & \textbf{0.7\%} & 13.6\% & 29.0\% & 42.2\% & 61.7\% & 63.2\% \\
\multirow{2}{*}{C4} & TPR & \textbf{90.3\%} & 65.8\% & 0.0\%  & 24.6\% & 22.6\% & 56.1\% \\
                    & FDR & 3.7\% & 11.8\% & \textbf{0.0\%}  & 2.4\% & 74.2\% & 12.1\% \\
\hline
\end{tabular}
}
{\scriptsize
\begin{tablenotes}
    \item(1) \textbf{MLP(5000)}: Additional test of MLP model trained with 5000 samples due to its higher data demand. All other non-AL models are trained with 500 samples.
    \item(2) \textbf{Higher TPR is preferred}: the surrogate model detects more transient instability events labelled by the benchmark TDS; \textbf{Lower FDR is preferred}: the surrogate model produces fewer false alarms for transient instability event prediction.
\end{tablenotes}
}
\vspace{-0.1cm}
\end{table}

Fig.~\ref{fig:240BusTrFPorb} demonstrates that the proposed AL-Kriging method consistently provides accurate predictions of instability probabilities. 
In comparison, AL-RF tends to underestimate the instability probability.
This is because the absence of unstable samples in the initial training dataset causes the entropy-based learning function to struggle to capture the low-probability instability region.


Moreover, Table~\ref{tab:tpr_fdr_240} highlights the superior pointwise detection accuracy of AL-Kriging for rare instability events, characterized by high TPR and low FDR across all contingencies,
which verifies its ability to detect potential real-life operating conditions of RES that could cause rare instability events. 
For example, the unstable samples for C1 are often characterized by both low wind generation at Bus 1034 and high wind generation at Bus 4035. 
On the other hand, although the non-AL methods occasionally yield accurate probability predictions (e.g., Kriging results for C3 in Fig.~\ref{fig:240BusTrFPorb}), their pointwise predictions suffer from either elevated false alarm rates (high FDR in  Table~\ref{tab:tpr_fdr_240}), incomplete detection of unstable samples (low TPR in Table~\ref{tab:tpr_fdr_240}), or both. Similar limitations are  observed for the deep learning model MLP trained with an order-of-magnitude larger dataset, highlighting the challenge
of rare-event identification without targeted sampling.
 


\subsubsection{Performance Comparison on Computational Cost:}


Table \ref{tab:ALK_Time_240Bus} presents the computational costs of the AL-Kriging and non-AL methods. For the non-AL methods, only the average cost is reported, as their cost is dominated by $t_{\mathrm{ed}}$, the execution time of time-domain simulations (TDS). 

\begin{table}[!]
\caption{Computation Cost of the AL-Kriging and Non-AL methods (240-Bus System)}
\label{tab:ALK_Time_240Bus}
\centering
\begin{tabular}{cccccc}
\hline
\makecell[c]{Method\\(Ctg.)} & \makecell[c]{AL-K\\(C1)} &  \makecell[c]{AL-K\\(C2)} & \makecell[c]{AL-K\\(C3)}  & \makecell[c]{AL-K\\(C4)} & \makecell[c]{Non-AL\\(Avg.)}  \\ \hline
$l_{\mathrm{total}}$ & $15$  & $34$  & $24$  & $27$ &  -  \\  
$N_{\mathrm{total}}$ & $190$  &  $380$  &  $280$ &  $310$ & $500$   \\  
$t_{\mathrm{ed}}$(s)   &  $1739.1$ & $3478.2$  &  $2562.9$ &  $2837.5$  & $4576.6$  \\  
$t_{\mathrm{sg}}$(s)  &  $29.7$ & $144.7$  &  $71.4$ &  $89.2$ & $15.9$ \\  
$t_{\mathrm{total}}$(s)  &  $1768.9$ & $3622.9$  &  $2634.3$ &  $2926.7$ & $4592.5$ \\  \hline
\end{tabular}
{\scriptsize
\begin{tablenotes}
\item * \textbf{AL-K (C1)}: the cost of AL-Kriging method on contingency C1. 
\item * \textbf{Non-AL (Avg.)}:  the average cost of the three non-AL methods over all contingencies.
\item **  Definition of symbols (e.g., $l_{\mathrm{total}}$): refer to Table \ref{tab:ALK_Time_59Bus}.
\end{tablenotes}
}
\end{table}

Table \ref{tab:ALK_Time_240Bus} shows that (i) the average processing time per TDS execution increases from $2.5$s (59-Bus system) to $9$s as system size grows. 
(ii) The proposed AL-Kriging maintains superior efficiency over the non-AL methods in detecting rare instability events.

Moreover, despite increases in system size and input dimensionality 
(from $23$ to $35$ inputs), the time required to build the AL-Kriging model, $t_{\mathrm{sg}}$, remains comparable to the 59-bus case 
(Table \ref{tab:ALK_Time_59Bus}). This is because the training cost of AL-Kriging is influenced more by the complexity of the transient instability domain 
$\mathcal{F}$ than by system size or input dimensions. 
Although difficult to quantify, a more complex instability domain may lead to a slower convergence rate of the active learning process and, consequently, a longer total computational time 
(e.g., C1 in Table \ref{tab:ALK_Time_59Bus}; C2 in Table \ref{tab:ALK_Time_240Bus})
and a lower pointwise detection accuracy for rare instability events  (e.g., the AL-Kriging results for C1 and C2 in Table \ref{tab:tpr_fdr_59}). 
\color{black}

\color{black}
\subsection{Analysis of Algorithm Parameters}
\label{subsec:ParameterAnalysis}

\subsubsection{Impact of Selection Pool Size:} 
\label{subsubsec:PoolSizeTest}

The size $N_V$ of the sample selection pool $\mathcal{X}_V$ for AL-Kriging should be large enough to ensure sufficient coverage of the rare instability region, that is, it should include a sufficient number of unstable samples for effective sample selection. Since this number is mainly determined by $P_f$, the required size of $\mathcal{X}_V$ is expected to depend more on the level of $P_f$ than on the system scale or input dimensionality.

To verify this point, a comparative study is conducted on the 59-bus and 240-bus systems. Contingency C3 in Table \ref{tab:59BusCtg} and contingency C1 in Table \ref{tab:240BusCtg} with an adjusted fault clearing time are selected. For each contingency, a sample selection pool $\mathcal{X}_V$ of size $10^3$ or $10^5$ is adopted. The resulting Kriging models are then compared with TDS-MCS results of $10^5$ samples to evaluate their classification accuracy.

The TPR and FDR results in Table \ref{tab:NV_impact} indicate that a candidate pool of $10^3$ samples is sufficient for the 240-bus system with $P_f\approx15\%$, but not for the 59-bus system with $P_f\approx1\%$. This implies that the required size of the sample selection pool is primarily determined by the level of $P_f$, rather than by the system scale or the number of uncertain inputs.



\begin{table}[!h]
\caption{Impact of the sample selection pool size $N_V$ on the performance of AL-Kriging}
\label{tab:NV_impact}
\centering
\begin{tabular}{l c c c c c c}
\toprule
Test case & \makecell[c]{No. of \\ inputs } & \makecell[c]{True $P_f$ } & $N_V$ & \makecell[c]{Predicted $P_f$ } & TPR & FDR \\
\midrule
\multirow{2}{*}{\makecell[l]{Ctg. C3 \\ 59-bus}}
& \multirow{2}{*}{23} 
& \multirow{2}{*}{1.23\%} 
& $10^3$ 
& 0.92\% 
& 67.2\% 
& 10.6\% \\
&  &
& $10^5$ 
& 1.18\%
& 90.3\%  
& 6.2\% \\
\midrule
\multirow{2}{*}{\makecell[l]{\textit{Modified} C1\\240-bus }}
& \multirow{2}{*}{35} 
& \multirow{2}{*}{14.5\%} 
& $10^3$ 
& 14.0\%
& 92.6\% 
& 4.4\% \\
&  &
& $10^5$ 
& 14.0\%
& 91.8\% 
& 5.0\% \\
\bottomrule
\end{tabular}
{\scriptsize
\begin{tablenotes}
    \item(1) \textbf{Modified Ctg. C1, 240-bus}: Contingency C1 in Table \ref{tab:240BusCtg} is modified to have a larger fault clearing time $t_{\mathrm{fct}}$ of 7 cycles, rendering a larger instability probability $P_f$ around 14\%.
    \item(2) \textbf{True \& Predicted $P_f$}: The true $P_f$ is estimated by TDS-MCS using $10^5$ samples. The predicted $P_f$ is evaluated on the same samples using AL-Kriging in place of TDS.
\end{tablenotes}
}
\end{table}

\subsubsection{Impact of Stopping Criterion and Iteration Limit:} 
\label{subsubsec:IterLimitTest}
\begin{figure}[h!]
    \centering    \includegraphics[width=0.7\columnwidth]{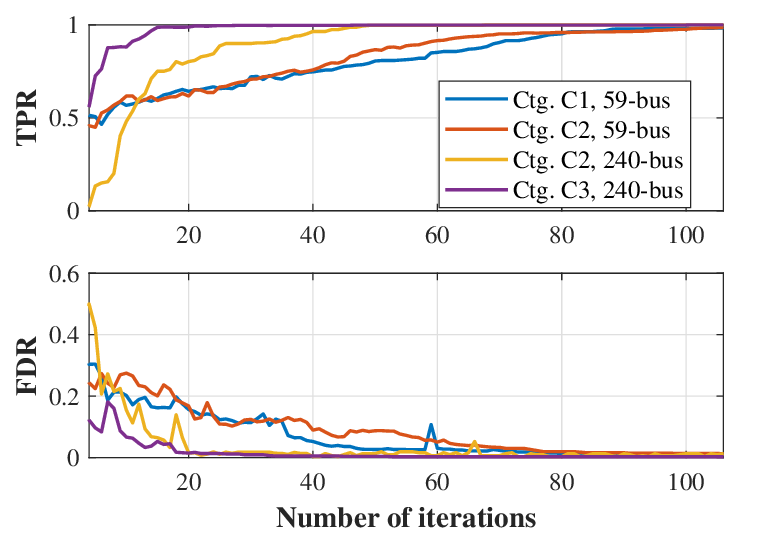}
    \caption{Comparison of the true positive rate (TPR) and false discovery rate (FDR) of the Kriging models during the active learning process for selected contingencies in the 59-bus and 240-bus systems. For this comparison, the stopping criterion is disabled, and the iteration limit is increased to $l_{max}=106$.}
    \label{fig:TPR_FDR_Convergency}
\vspace{-0.4cm}
\end{figure}

The active learning (AL) process terminates when the stopping criterion is satisfied or the iteration limit $l_{max}$ is reached. These two mechanisms alleviate the computational burden by limiting excessive sample enrichment. However, they may also reduce the accuracy of AL-Kriging when the AL process converges slowly. This effect can be observed for contingencies C1 and C2 in Table \ref{tab:tpr_fdr_59}, which exhibit relatively poor unstable sample detection performance. To further illustrate the impact of these two mechanisms, the stopping criterion is disabled and the iteration limit is increased to $l_{max}=106$. Fig. \ref{fig:TPR_FDR_Convergency} shows the resulting TPR and FDR trajectories over the AL iterations for selected contingencies with $P_f<1\%$.

Fig. \ref{fig:TPR_FDR_Convergency} shows that AL-Kriging converges slowly under contingencies C1 and C2 of the 59-bus system. In these cases, the AL process requires around 70 iterations to achieve satisfactory unstable sample detection accuracy (TPR > 90\% and FDR < 5\%). This slow convergence explains the relatively poor accuracy reported in Table \ref{tab:tpr_fdr_59} under the preset iteration limit of $l_{max}=40$. In contrast, for the larger-scale 240-bus system with more inputs, AL-Kriging can achieve satisfactory detection accuracy within 40 iterations. This suggests that the slow convergence is less related to the system scale or the input dimensionality and is more likely caused by the complexity of the stability boundary in the input space.

\subsubsection{Impact of Sample Enrichment Size Per Iteration:}
\label{subsubsec:Ne_Test}

\begin{figure}[h!]
    \centering    \includegraphics[width=0.7\columnwidth]{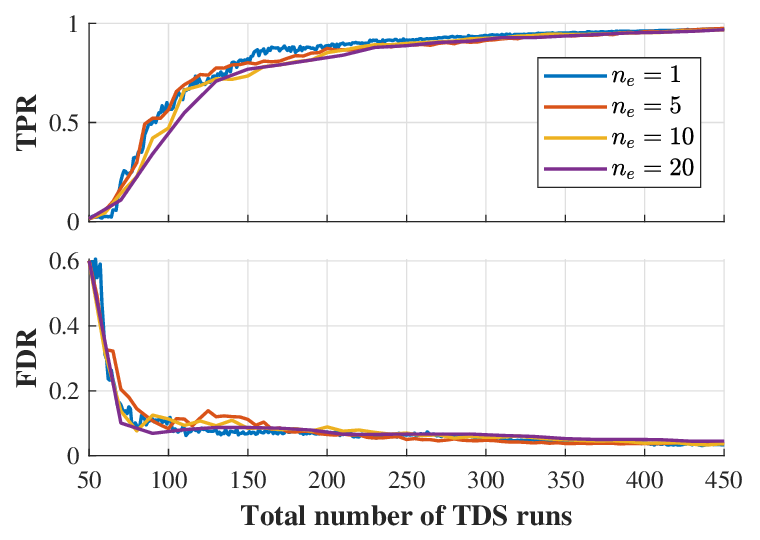}
    \caption{Comparison of the true positive rate (TPR) and false discovery rate (FDR) of AL-Kriging with different numbers of enriched TDS runs per iteration, $n_e$. Contingency C3 of the 59-bus system is used as an example.}
    \label{fig:ne_Convergency}
\vspace{-0.4cm}
\end{figure}

During the active learning (AL) process, we perform $n_e$ times of time-domain simulations (TDS) per iteration to enrich the Kriging training dataset. Under the same total number of TDS runs, a larger $n_e$ leads to fewer Kriging retraining steps, thereby reducing the computational burden caused by repetitive model training. However, it also results in fewer updates of the learning function-based sampling strategy, which may lead to slower convergence of AL-Kriging. To examine this trade-off, we compare the unstable sample detection accuracy of AL-Kriging under different $n_e$ values as the total number of TDS runs increases. The results are shown in Fig. \ref{fig:ne_Convergency}.

Fig.~\ref{fig:ne_Convergency} indicates that $n_e$ has only a minor impact on the convergence of AL-Kriging. Although only contingency C3 of the 59-bus system is presented here for illustration, the same observation also holds for the other contingencies. Therefore, a reasonably large $n_e$ (e.g., $n_e=10$) can be adopted to reduce the Kriging training burden in AL-Kriging. However, an excessively large $n_e$ (e.g., $n_e=50$) is not recommended, as the stopping criterion requires a series of $P_f$ predictions to assess the convergence behavior.
\color{black}

\section{Conclusions}
\label{sec:conclusion}

This study presents a Kriging-based active learning framework (AL-Kriging) that enables efficient probabilistic assessment of transient instability events with small occurrence probabilities under load and RES uncertainties.

The proposed framework is validated on both a modified IEEE 59-bus system and a large-scale WECC 240-bus system incorporating real-world renewable generation data.
Compared with the existing AL-RF method and three non-AL methods, the proposed AL-Kriging consistently provides
more reliable identification of rare transient instability events using a limited number of time-domain simulations.
These findings highlight the necessity of explicitly addressing rare and extreme tail events for accurate probabilistic transient stability assessment, as existing methods, even when trained with a much larger dataset, may misidentify critical low-probability instability regions that are relevant to system resilience analysis.
Moreover, the proposed AL-Kriging framework offers insights into the critical operating conditions of RES and loads associated with rare transient instability events, which can support resilience-oriented planning and operation.

Future work includes improving  prediction accuracy for complex instability domains and leveraging AL-Kriging
to precisely locate instability regions, thereby enabling sensitivity analyses for rare transient instability events.

\section*{Acknowledgements}
This work was supported by Natural Sciences and Engineering Research Council (NSERC) Discovery Grant, NSERC RGPIN-2022-03236, CRC-2023-00006,  and by Fonds de recherche du Québec under Grant FRQ-NT 2023-NOVA-314338.

\bibliographystyle{elsarticle-num}
\bibliography{PTSA.bib}

\end{document}